\documentclass[lettersize,journal]{IEEEtran}
\usepackage{amsmath,amsfonts}
\usepackage{algorithmic}
\usepackage{array}
\usepackage[caption=false,font=normalsize,labelfont=sf,textfont=sf]{subfig}
\usepackage{textcomp}
\usepackage{stfloats}
\usepackage{url}
\usepackage{verbatim}
\usepackage{graphicx}
\def\BibTeX{{\rm B\kern-.05em{\sc i\kern-.025em b}\kern-.08em
    T\kern-.1667em\lower.7ex\hbox{E}\kern-.125emX}}
\usepackage{balance}

\usepackage[T1]{fontenc}% optional T1 font encoding
\usepackage{amsmath}
\usepackage{listings}
\usepackage{amsmath,amssymb,amsfonts}
\usepackage[ruled,vlined]{algorithm2e}
\usepackage{multirow}
\usepackage{textcomp}
\usepackage{enumerate}
\usepackage{xcolor}
\usepackage{diagbox}
\usepackage{amsthm}

\usepackage{resizegather}
\usepackage{subfloat}
\usepackage[numbers,sort&compress]{natbib}
\graphicspath{{figures/}}

\begin{document}
\title{A Novel Reversible Data Hiding Scheme Based
	on Asymmetric Numeral Systems}
\author{
	Na~Wang,~\IEEEmembership{Member,~IEEE,}
	Chuan~Qin,~\IEEEmembership{Member,~IEEE,}
	Sian-Jheng~Lin,~\IEEEmembership{Member,~IEEE}
	
	\IEEEcompsocitemizethanks{\IEEEcompsocthanksitem Na~Wang and Chuan~Qin are with the School of Optoelectronic Information and Computer Engineering, University of Shanghai for Science and Technology, Shanghai 200093, China. \protect
	E-mail: wna@usst.edu.cn; qin@usst.edu.cn.
	\IEEEcompsocthanksitem Sian-Jheng~Lin is with the Theory Laboratory, Huawei Technologies Company Ltd., Hong Kong, China. \protect
	E-mail: lin.sian.jheng1@huawei.com.
	}
}

\markboth{Journal of \LaTeX\ Class Files,~Vol.~18, No.~9, September~2020}%
{How to Use the IEEEtran \LaTeX \ Templates}

\maketitle

\begin{abstract}
Reversible data hiding (RDH) has been extensively studied in the field of information security. In our previous work~\cite{6194314}, an explicit implementation approaching the rate-distortion bound of RDH has been proposed. However, there are two challenges left in our previous method. Firstly, this method suffers from computing precision problem due to the use of arithmetic coding, which may cause the further embedding impossible. Secondly, it had to transmit the probability distribution of the host signals during the embedding/extraction process, yielding quite additional overhead and application limitations. In this paper, we first propose an RDH scheme that employs our recent asymmetric numeral systems (ANS) variant as the underlying coding framework to avoid the computing precision problem. Then, we give a dynamic implementation that does not require transmitting the host distribution in advance. The simulation results show that the proposed static method provides slightly higher peak signal-to-noise ratio (PSNR) values than our previous work, and larger embedding capacity than some state-of-the-art methods on gray-scale images. In addition, the proposed dynamic method totally saves the explicit transmission of the host distribution and achieve data embedding at the cost of a small image quality loss.
\end{abstract}


\section{Introduction}
\IEEEPARstart{R}{eversible} data hiding (RDH) aims at secure transmission of secret data through communication media. This is achieved through embedding of secret data into the host media, and it has the property that the receiver can losslessly reconstruct the host media and secret data from the received steganographic (abbreviated as stego) media. A media used to embed data is defined as the host media, including image, text file, audio, video and multimedia, and the media that has embedded secret data is called the stego media. 
RDH is widely adopted in medical imagery~\cite{gao2021reversible}, image authentication~\cite{4453842} and law forensics, where the host media is so precious that cannot be damaged.

In the past two decades, many RDH works have emerged and they are categorized mainly into plain domain and encrypted domain. Plain domain approaches of RDH are designed to achieve high embedding capacity and low distortion without any protection of the host media, where embedding capacity is defined as a ratio between the number of secret data bits and host signals. The purpose of RDH in encrypted domain is to embed additional data into the encrypted media without revealing the plaintext data.
This paper focuses on the RDH methods in the plain domain. Popular plain domain works roughly categorized as i) difference expansion (DE)~\cite{5762603,1227616,4703233,9429907}, ii) histogram shifting (HS)~\cite{4036851,7393820,6459018,8709802} and iii) lossless compression~\cite{1381493,6497612,lin2015novel,9653806}. In the DE-based RDH methods, the differences of each pixel group are expanded, e.g., multiplied by $2$, so that the lowest significant bits (LSB) of the differences are all zero and can be used to embed data. In \cite{1227616}, Tian introduced the first difference expansion method based on integer transform, but it suffers from low embedding capacity and its overhead cost was almost double the amount of hiding data. To remedy this problem, A.M. Alattar~\cite{1315703} gave a generalized integer transform, and pixel block of arbitrary
size was considered instead of pixel pairs, to improve the embedding capacity. However, this method is applicable only for color images. 

Histogram is a graphical representation of data points. The maximum and minimum number of occurrence of elements in the set of data points are called peak point and zero point, respectively. Histogram shifting (HS) is an effective method for RDH, it improves the embedding capacity by enlarging the peak pixels. When images are selected as the host media, the peak signal-to-noise ratio (PSNR) is often used as a metric to measure the performance of an RDH scheme, and the higher the PSNR, the better the stego image quality. In 2006, Ni et al.~\cite{1608163} chosen the bins with the highest frequencies in the histogram to embed the secret bits.  
In this method, each pixel value is modified at most by 1, thus the PSNR of stego image is guaranteed. Yet, this approach suffers from the issues of multiple zero points, which requires the storage for the coordinate of zero points. In \cite{fallahpour2007high}, Fallahpour et al. improved the embedding capacity of Ni et al.'s method by applying HS on blocks instead of the whole image. However, this technique bears zero and peak point information that needs to be transmitted to the receiver for data retrieval. Tai et al.~\cite{4801625} presented an RDH scheme based on histogram modification. In this method, the binary tree structure was exploited to solve the issue of multiple pairs of peak points for communication. But, its pure embedding capacity unexpectedly decreases with increasing tree level and the overhead information increases as well. In~\cite{8733828}, Ou et al. proposed an RDH method based on multiple histogram modification to achieve high embedding capacity, by using multiple pairs of bins for every prediction error histogram. However,  the location map as overhead information is necessary to preserve reversibility, and it can be large even after compression.

The lossless compression algorithm is extensively explored in RDH technique. The main idea is to save space
for hiding additional data by performing lossless compression
on the host media. Celik~\cite{1381493} introduced a
generalization least significant bit (G-LSB) method to
enhance the compression effectiveness by using arithmetic
coding. In this method, the embedding capacity was enhanced as higher
level pixel was considered. However, when the higher level was used, the distortion of host media also increased. In \cite{manikandan2022high}, Manikandan employed run-length encoding and a modified Elias gamma encoding to embed some additional bits during the encoding process itself. Moreover, one natural problem is what is the upper limit of the embedding capacity for a given host media and distortion constraint. This problem was solved by Kalker et al.~\cite{1027818} for independent and identically distributed (i.i.d.) host sequence. In \cite{1027818}, Kalker and Willems formulated the rate-distortion function, i.e., the upper bound $R_{max}$ of the embedding rate under a given distortion constraint $\Delta$ as follows:
\begin{equation}\label{eq.bound}
	R_{max}=\text{maximize} \left\{H(Y) \right\} -H(X),
\end{equation}
where $X$ and $Y$ denote the random variables of host media
and stego media respectively, and $H(\cdot)$ is the Shannon entropy of a discrete random variable. The entropy
is maximized over all transition probabilities $P_{Y|X}(y|s)$ satisfying
the distortion constraint 
\begin{equation}
	\sum_{s\in X, y\in Y}{P_X(s)P_{Y|X}(y|s)D(s,y)}\leq \Delta,
\end{equation}
where $P_X(s)$ is the probability mass function (pmf) of host sequence $X$, and the distortion metric $D(s,y)$ is usually defined as the square
error distortion, i.e., $D(s,y)=(s-y)^2$. Therefore, to evaluate the embedding capacity of an RDH scheme, the optimal transition probability (OTP) $P_{Y|X}(y|s)$ needs to be calculated. For some specific distortion metrics $D(s,y)$, such as the square error distortion $D(s,y)=(s-y)^2$ or L1-Norm $D(s,y)=|s-y|$, it has been proved that the OTP has a Non-Crossing-Edges (NCE) property~\cite{01}. Based on the NCE property, our previous work~\cite{6194314} found that the optimal solution on $P_{Y|X}(y|s)$ can be derived from the analysis of pmfs $P_X(s)$ and $P_Y(y)$.
Because $P_X(s)$ is available from the given host sequence $X$, this means that the encoder and decoder only need to calculate the $P_Y(y)$ for the distortion metric that satisfies the NCE property. 

In our previous work~\cite{6194314}, we proposed a backward and forward iterative (BFI) algorithm to estimate the optimal pmf $P_Y(y)$. Also, we gave an arithmetic coding \cite{5391119}-based
scheme that can approach the rate-distortion bound~\eqref{eq.bound}. After that, Hu et al.~\cite{6493430} introduced a fast algorithm to estimate the $P_Y(y)$ in our previous work. However, there are still two challenges unresolved in our previous work~\cite{6194314}. Firstly, we adopted arithmetic coding as the underlying coding framework in \cite{6194314} to embed a message in the i.i.d. host sequence. In the arithmetic coding, there is a limit to the precision of the number that can be encoded, which constrains the number of symbols to encode within a codeword. As a result, our previous work~\cite{6194314} suffers from the computing precision problem in some special cases. For example, if the coding interval in the arithmetic coding is narrowed to zero, then further data embedding is impossible. Secondly, both the embedding and extraction processes of
methods \cite{6194314}, \cite{6493430} require transmitting the host pmf $P_X(s)$ as parameters (for the purpose of estimating $P_Y(y)$), which costs quite additional overhead, and the overhead size is proportional to the alphabet size. This is usually unacceptable for extremely memory-constrained scenarios, and it has certain limitations in practical applications.

To fully address the above two challenges left in our previous work~\cite{6194314}, we first propose a code construction scheme that employs our recent asymmetric numeral systems (ANS) variant~\cite{9810728}. ANS~\cite{duda2009asymmetric,7170048,9478894,8849430} is a family of entropy coding introduced by Jarek Duda in 2009. It has been popular and incorporated into many compressors, including Facebook Zstandard, Apple LZFSE and Google Draco 3D, because it provides the compression ratio of arithmetic coding with a processing speed similar to that of Huffman coding~\cite{4051119}.
To the best of our knowledge, this paper is the first that employs ANS coding for RDH technique. More importantly, unlike arithmetic coding, ANS coding is not subject to the computing precision. Therefore, by using ANS coding as the underlying coding framework, our proposal can avoid the precision problem that may occur in our previous work~\cite{6194314}. Second, we give a dynamic embedding/extraction implementation that does not require to explicitly transmit the host pmf, thus saving additional overhead costs.

This paper considers the host media composed of i.i.d. gray-scale samples. The contributions of this paper are summarized as follows.
\begin{enumerate}
	\item A code construction scheme that employs our recent ANS variant is proposed to address the computing precision problem in our previous work~\cite{6194314}. To the best of our knowledge, this paper is the first that employs ANS coding for RDH technique.
	\item A dynamic implementation is given that does not require prior transmission of the host pmf during embedding and extraction.
	\item The simulations are given.
	\begin{enumerate}
		\item For the host sequences drawn from discrete normal distribution: the proposed static method provides similar embedding performance to our previous work; the proposed dynamic implementation saves the additional overhead of explicitly transmitting the host pmf at the cost of a slight embedding distortion.
		\item For the gray-scale images: the proposed static method provides slightly higher PSNR values than our previous work and larger embedding rates than some state-of-the-art RDH methods; the proposed dynamic method employs predicted pmf instead of true pmf to save additional overhead and produces a small image quality loss.
	\end{enumerate}
\end{enumerate}

The rest of this paper is organized as follows. Section \ref{sec:2} lists the notations and related works. Section \ref{sec:3} introduces the proposed method to achieve the rate-distortion bound. In Section \ref{sec:4}, we give a dynamic embedding/extraction implementation without the need to transmit the host pmf. Section \ref{sec:5} shows the simulation results and Section \ref{sec:6} concludes this work.

\section{Preliminaries}\label{sec:2}
\subsection{Notations}\label{sec:2.1}
Throughout this paper, the
host sequence $X=\left(x_1, x_2, \cdots, x_N\right)$ is composed of i.i.d. samples drawn from the probability mass function (pmf) $P_X=\left\{P_X(s)| s\in \Sigma_B\right\}$, where the set $\Sigma_B=\left\{0, 1, \cdots, B-1\right\}$ is an alphabet of size $B$. Let $P_{CX}=\left\{P_{CX}(s)=\sum_{i=0}^{s}P_X(i)|s\in \Sigma_B \right\}$ denote the cumulative pmf of $X$. It is noted that we have $P_{CX}(-1)=0$ and $P_{CX}(B-1)=1$. Let $f_X(s)$ denote the number of occurrences of signal $s$ in the sequence $X$, and $c_X(s)\triangleq \sum_{i=-1}^{s-1}f_X(i)$ is the cumulative frequency of signal $s$, where $f_X(-1)=0$ by default.
The entropy function $H(X)$ of a discrete random variable $X$ is defined as $H(X)=-\sum_{i=0}^{B-1}P_X(i)\log_2{P_X(i)}$. The random variable $W$ represents a message uniformly distributed in $\Sigma_M=\left\{0, 1, \cdots, M-1\right\}$. This paper considers embedding binary secret data into gray-scale signals, so we have $B=256$ and $M=2$.

Given a binary stream $S$, we construct a LIFO (Last In First Out) stack accordingly. The operation
\begin{equation}\nonumber
	Q \gets Read_S(i)
\end{equation}
pops $i$ bits $\left\{S_i\right\}_{i=0}^{i-1}$ from the bottommost position in the stack that contains $S$, and these $i$ bits form an integer $Q=S_0\times 2^{i-1}+\cdots+S_{i-2}\times 2+S_{i-1}$. The operation $Write_S(Q,i)$ pushes the least significant $i$ bits of an integer $Q$ to the stack.

\begin{figure}[t]
	\centering
	\includegraphics[width=3.7in]{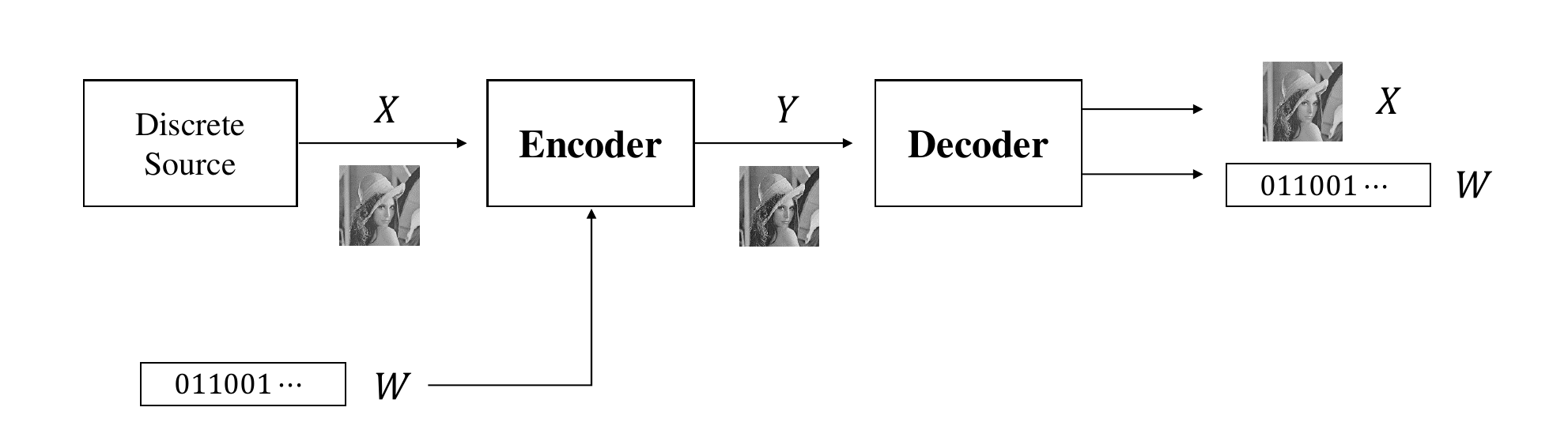}
	\caption{An example of RDH on a gray-scale image.}  
	\label{fig.1}
\end{figure}

In RDH, $L$ bits of message $W=\left(w_1, w_2, \cdots, w_L\right)$ are embedded by the encoder into the host sequence $X=\left(x_1, x_2, \cdots, x_N\right)$, by
slightly modifying its elements to produce a stego sequence $Y=\left(y_1, y_2, \cdots, y_N\right)$, where $w_i\in \Sigma_M$ and $x_j\in \Sigma_B$, $y_j\in \Sigma_Z$ for $1\leq i\leq L$, $1\leq j\leq N$. This paper considers the convention that host signals and stego signals draw from the same set, i.e., $\Sigma_Z=\Sigma_B$.
We denote the embedding capacity, also called embedding rate, by $R=L/N$. Schemes are usually constructed to minimize some distortion measure $D(s,y)$ between the host sequence $X$ and the stego sequence $Y$ for a given embedding rate $R$. The distortion metric $D(s,y)$
in this paper is defined as the square error
distortion, i.e., $D(s,y)=(s-y)^2$. The decoder can losslessly reconstruct the message $W$ and the host sequence $X$ from the stego sequence $Y$. Figure \ref{fig.1} shows the general framework of RDH on a gray-scale image $X$, where $Y$ is the stego image with some distortion after embedding $W$.

\subsection{Related works}\label{sec:2.2}
\subsubsection{A Near Optimal Coding Method using Arithmetic Coding}\label{sec:2.2.1}
In our previous work \cite{6194314}, we proposed a backward and forward iterative (BFI) algorithm to estimate the optimal distribution $P_Y$ of
the stego signals. The BFI algorithm assumes that the
host and stego signals draw from the same set, i.e., $\Sigma_Z=\Sigma_B$. Algorithm \ref{BFI} gives the details. Notably, the BFI method controls an input parameter $\alpha$ for various embedding rate and distortion pairs. The case $\alpha=1$ admits the uniform distribution $P_Y=\left\{P_Y(y)=1/B| y\in\Sigma_B \right\}$ performing the maximal embedding rate, and another case $\alpha=+\infty$ generates an unchanged distribution $P_Y=P_X$ with zero embedding rate.

\begin{algorithm}[t]
	\caption{\label{BFI} The BFI algorithm \cite{6194314} to estimate the optimal cumulative pmf $P_{CY}$}
	\LinesNumbered
	\KwIn{The cumulative pmf $P_{CX}$ of the host sequence $X$, a real
number $\alpha$ and the tolerance $\varepsilon$.}
	\KwOut{The cumulative pmf $P_{CY}$ of the stego sequence $Y$.}
    Given an initial set $E=\left\{e_y=(y+1)/B, y= -1 \;\text{to} \;B-1 \right\}$, or the user can design an arbitrary initialization as long as $0=e_{-1}\leq e_0 \leq \cdots \leq e_{B-1}=1$. Declare the variable $var=0$.\\
    For $y$ from $0$ to $B-2$, update each $e_y$ through
        $$e_y^{new} = \begin{cases}
    	\frac{e_{y+1}-e_{y-1}}{1+\alpha^{D(s,y)-D(s,y+1)}} &\text{if there exists}\; e_y \in\\ +e_{y-1}, &  (P_{CX}(s-1), P_{CX}(s));\\
    	&\text{if there exists} \\
   	    &\alpha^{D(s,y)-D(s,y+1)}\\
    	P_{CX}(s), &\leq \frac{e_{y+1}-P_{CX}(s)}{P_{CX}(s)-e_{y-1}}\\
    	&\leq \alpha^{D(s+1,y)-D(s+1,y+1)};
    \end{cases}$$\\
After finding the new value $e_y^{new}$, record the maximal offset by 
$$var= max\left\{var, |e_y^{new}-e_y^{old}|\right\}.$$\\
For $y$ from $B-2$ to $0$, update each $e_y$ and the maximal offset $var$ through the same criterion in steps $2$--$3$.\\
If $var\geq \varepsilon$, set $var=0$ and then go to step 2; otherwise, output $P_{CY}=\left\{P_{CY}(y)=e_y\right\}$.
\end{algorithm}

Based on the result of the BFI algorithm, we proposed a coding scheme in \cite{6194314} to embed a message $W$ into the host sequence $X=\left(x_1, x_2, \cdots, x_N\right)$ by the given cumulative pmfs $P_{CX}$ and $P_{CY}$. The coding framework is similar to arithmetic coding. The encoder maintains two values $\left(l^{(i)},u^{(i)}\right) \subseteq \left[0,1\right]$, $i=1,2, \cdots, N$ to interpret the joint information of the host signal $x_i$ and the temporal information produced at the last step. At the initialization, the vector $\left(l^{(1)},u^{(1)}\right)=\left(P_X(x_1-1), P_X(x_1)\right)$ represents the information of the first host signal. At the $i$-th step, the encoder transforms the vector $\left(l^{(i)},u^{(i)}\right)$ and the next host $x_{i+1}$ to the stego signal $y_i$ and the next vector $\left(l^{(i+1)},u^{(i+1)}\right)$, expressed as $\left(y_i,l^{(i+1)},u^{(i+1)} \right)=Encode\left(x_{i+1},l^{(i)},u^{(i)} \right)$. First, we
need to find the possible $y_i$ within the interval $(l^{(i)},u^{(i)}]$, then $y_i \in \left[y_1^{(i)},y_2^{(i)}\right]$ if and only if $[P_Y(y_1^{(i)}),P_Y(y_2^{(i)}-1)) \subset [l^{(i)},u^{(i)}) \subseteq [P_Y(y_1^{(i)}-1),P_Y(y_2^{(i)}))$. Second, for the case $y_1^{(i)}=y_2^{(i)}$, the stego signal $y_i=y_1^{(i)}$ is determined. The $y_1^{(i)}$ determines the interval $[P_Y(y_1^{(i)}-1),P_Y(y_1^{(i)}))$, which is larger than $(l^{(i)},u^{(i)}]$, so for lossless recovery, the decoder needs more information, expressed as the information vector $\left(P_Y(y_1^{(i)}-1), l^{(i)},u^{(i)}, P_Y(y_1^{(i)})\right)$ for lossless decoding. One can see that in order to determine $y_i$, the pmf $P_Y$ needs to be known first, and it can be derived from the cumulative pmf $P_{CY}$, i.e., $P_Y=\left\{P_Y(y)=P_{CY}(y)-P_{CY}(y-1)| y\in \Sigma_B\right\}$.
The next host signal $x_{i+1}$ determines the interval $[P_X(x_{i+1}-1),P_X(x_{i+1}))$, and we scale the information vector to $[P_X(x_{i+1}-1),P_X(x_{i+1}))$ to obtain $\left(l^{(i+1)},u^{(i+1)} \right)=\left(F(l^{(i)},x_{i+1},y_i),F(u^{(i)},x_{i+1},y_i) \right)$, where 
\begin{equation}
	F(k,s,y)=\frac{P_X(s)-P_X(s-1)}{P_Y(y)-P_Y(y-1)}\left(k-P_Y(y-1)\right)+P_X(s-1),
\end{equation}
for $k\in \mathbb{R}$, $s,y\in \Sigma_B$.
Then, we proceed to process the next host signal.

For another case $y_1^{(i)}>y_2^{(i)}$, the $y_i$ has several possible values corresponding to each interval of $$\left(l^{(i)}, P_Y(y_1^{(i)}), \cdots, P_Y(y_2^{(i)}-1), u^{(i)} \right).$$ The adaptive
arithmetic decoding is applied on the binary representation of message $W$ to determine the value $y_i$. To fit the usage requirement of arithmetic coding, the two ends of the vector are scaled to $0$ and $1$, resulting in $\left(0, G(P_Y(y_1^{(i)}),l^{(i)},u^{(i)}), \cdots, G(P_Y(y_2^{(i)}-1),l^{(i)},u^{(i)}), 1\right)$, where 
\[
G(k,l,u)=\frac{k-l}{u-1}, \qquad \text{for} \; k,l,u\in \mathbb{R}.
\]
The arithmetic decoder determines the value $y_i$, and the corresponding residual information that depends on the $y_i$ is
discussed below. For the case $y_1^{(i)}+1\leq y_i\leq y_2^{(i)}-1$, the $[P_Y(y_i-1),P_Y(y_i))$ is within the interval  $(l^{(i)},u^{(i)}]$, so we don't have the residual information, and the updated vector is $\left(l^{(i+1)},u^{(i+1)}\right)=(P_X(x_{i+1}-1),P_X(x_{i+1}))$ representing the next host signal. For the case $y_i=y_1^{(i)}$, the residual information is interpreted as the vector $(P_Y(y_i-1),l^{(i)},P_Y(y_i),P_Y(y_i))$, which is scaled into the interval $[P_X(x_{i+1}-1),P_X(x_{i+1}))$ to obtain the updated vector $\left(l^{(i+1)},u^{(i+1)}\right)=\left(F(l^{(i)},x_{i+1},y_i),P_X(x_{i+1})\right)$. For the case $y_i=y_2^{(i)}$, the residual information is interpreted as the vector $(P_Y(y_i-1),P_Y(y_i-1),u^{(i)},P_Y(y_i))$, which is scaled into the interval $[P_X(x_{i+1}-1),P_X(x_{i+1}))$ to obtain the updated vector $\left(l^{(i+1)},u^{(i+1)}\right)=\left(P_X(x_{i+1}-1),F(u^{(i)},x_{i+1},y_i)\right)$. In summary, the updated vector $\left(l^{(i+1)},u^{(i+1)}\right)$ is as follows:
\[
l^{(i+1)}=max\left\{P_X(x_{i+1}-1), F(l^{(i)},x_{i+1},y_i)  \right\}
\]
and
\[
u^{(i+1)}=min\left\{P_X(x_{i+1}), F(u^{(i)},x_{i+1},y_i)  \right\}.
\]
We continue the above steps until there is no next host signal. When $y_N$ is designated, the encoder sends out a prefix-free code value $v \in [\tilde{N}(l^{(N)},y_N), \tilde{N}(u^{(N)},y_N))$ with minimal code length, where
\[
\tilde{N}(s,y)=\frac{s-P_Y(y-1)}{P_Y(y)-P_Y(y-1)}.
\]
More details can be found in \cite{6194314}.

There are two observations from the above description. First, the practical implementation cannot store the real interval $\left(l^{(i)},u^{(i)}\right)$ due to the machine precision. Therefore, our previous work maintained the variables $\overline{L}=\left(\lfloor l^{(i)} \rfloor+1\right)\times 2^b$ and $\overline{H}=\lfloor u^{(i)} \rfloor\times 2^b$ with length $b$, where $\lfloor \cdot \rfloor$ returns the greatest integer less than or equal to the input number, and $b \in \mathbb{N}^+$. As the arithmetic coding proceeds, the $\overline{L}$ gradually approaches the $\overline{H}$ and the two values may coincide in certain cases. When $\overline{L}$ and $\overline{H}$ coincide, it is impossible to embed any data. That is, the usage of arithmetic coding here may suffer from computing precision problem. Second, it is necessary to transmit the cumulative pmfs $P_{CX}$ and $P_{CY}$, or only transmit $P_{CX}$ and use the BFI algorithm to generate $P_{CY}$. No matter which strategy is chosen, it costs an additional overhead proportional to the alphabet size.

\subsubsection{A Variant of ANS Coding (ANS-Variant)~\cite{9810728}}\label{sec:2.2.2}
The key idea of the pure form ANS coding is to encode a sequence of input symbols into a single natural number (also called state), unlike arithmetic coding that operates on a fraction. Before encoding, ANS holds a start state $x$. On a high level, the ANS encoder always maintains a state that can be modified by the encoding function $x'=\mathcal{C}(s,x)\in \mathbb{N}$: $\mathcal{C}(s,x)$ takes a symbol $s$ and the current state $x$ and produces a new state $x'$ that encodes both the symbol $s$ and current state $x$. Correspondingly, the decoding function $\mathcal{D}(x')$ decodes the state $x'$ to a symbol $s$ and the previous state $x$, i.e., $\mathcal{D}(x')=(s,x)$. It is noted that the encoder and decoder in ANS coding go through the same sequence of states, just in opposite order. This paper uses the convention that the encoder processes symbols forward, from the first to the last, and the decoder recovers symbols backwards, from the last to the first.

Consider an input sequence $X=\left(x_1, x_2, \cdots, x_N\right)$, it is assumed that $N=\sum_{s\in \Sigma_B}f_X(s)$ is a power of two, i.e. $N=2^n$ for an integer $n\geq 0$. This is a standard assumption like arithmetic coding for ANS coding to simplify arithmetic operations.
In our recent work \cite{9810728}, we proposed a new ANS variant that forces the state $x$ always at a fixed interval $I:=\left [2^{T-v\times n},2^T \right )$ by renormalization, where $T,v\in\mathbb{N}^+$ as long as $T-v\times n\geq 0$. Let $I_s:=\left [ f_X(s)\times 2^{T-v\times n},2^T \right )$ denote the interval corresponding to the symbol $s$, which ensures that the new state after encoding symbol $s$ is within the interval $I$. That is, if $x\in I_s$ we have $\mathcal{C}\left(x,s\right) \in I$; otherwise, we have $\mathcal{C}\left(x,s\right)\notin I$. The encoding function $\mathcal{C}\left(x,s\right)$ is defined as $\mathcal{C}\left(x,s\right)=\lfloor x/f_X(s) \rfloor$. For the ease of understanding, the encoding procedure is assumed to be implemented using a stack, where each entry in the stack is a bit. Below gives the encoding details. To encode each symbol $s$, if the current state $x$ is within $I_s$, we first push $n$-bit value $c_X(s)+(x\bmod {f_X(s)})$ to the stack and produce a new state $x$ by $x\gets \mathcal{C}\left(x,s\right)=\lfloor x/f_X(s) \rfloor$, where $\bmod$ denotes the remainder operation. Otherwise, we first pop $v\times n$ bits from the stack, and append them to the least significant $v\times n$ bits of $x$ to increase it. That is, if the decimal representation of the popped $v\times n$ bits is $D$, we have $x\gets (x\ll v\times n)+D$. After that, we push $n$-bit value $c_X(s)+(x\bmod {f_X(s)})$ to the stack and perform $\mathcal{C}\left(x,s\right)$ with the increased state $x$. Notably, to ensure that there are enough bits in the stack before the first popping, we initialize the start state to be large enough, e.g., $x = 2^T-1$. We follow the above steps until all symbols are encoded. Finally, it is necessary to transmit the final state of encoding (which is also the start state of decoding) and the remaining bits in the stack to the decoder for lossless recovery.

\begin{figure}[t]
	\centering
	\includegraphics[width=1.0\linewidth]{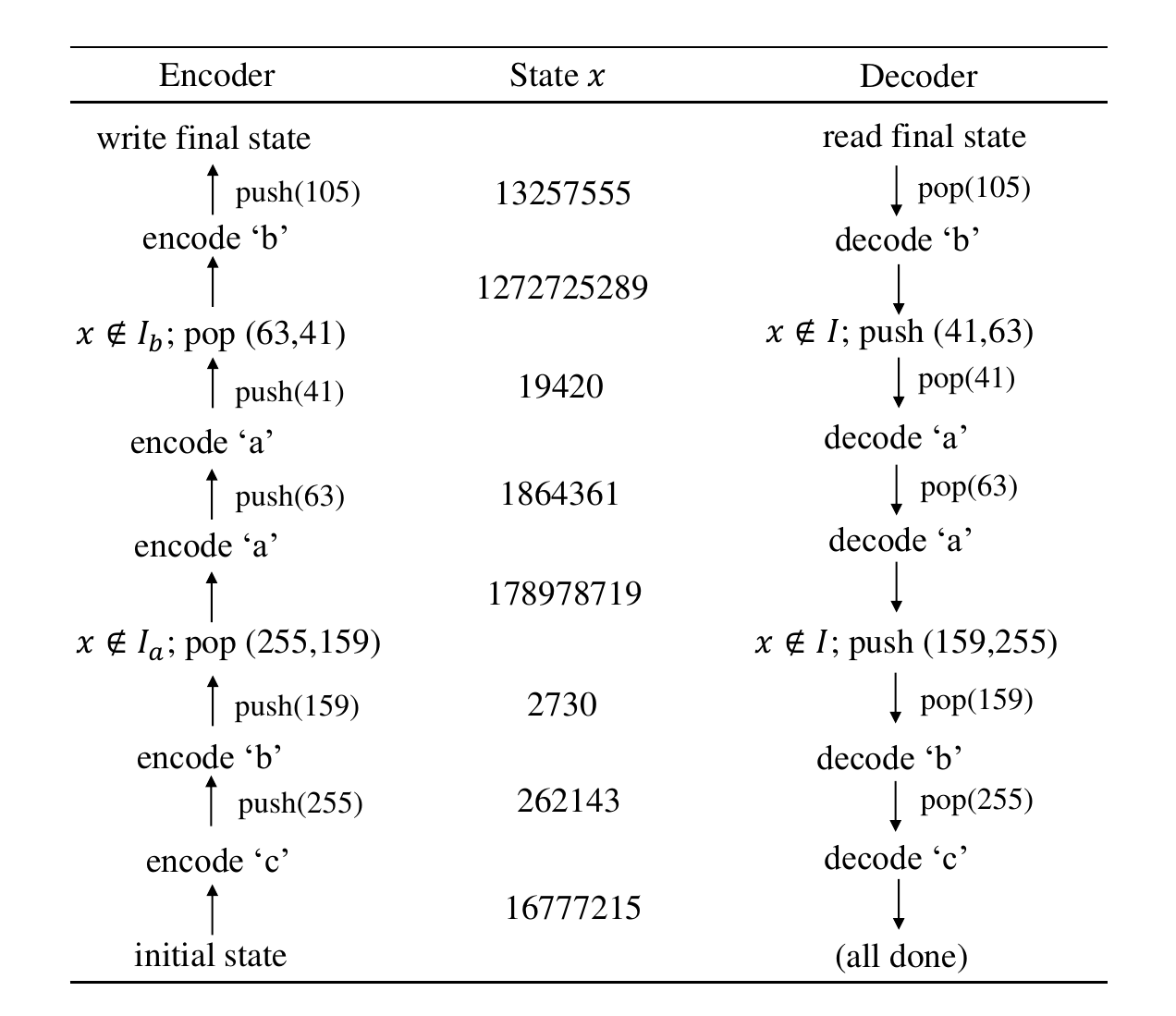}
	\caption{{Encoding example: coding a sequence $X=baabc$ with $P_X(a) =P_X(b)=\frac{96}{256}$, $P_X(c) = \frac{64}{256}$.
			The encoder proceeds from bottom to top, and the decoder proceeds from top to bottom (as indicated). Both go through the exact same sequence of states and perform I/O in the same places.}}
	\label{fig.3}
\end{figure}

Next, we discuss the corresponding decoding details of our recent ANS-Variant coding. The decoding procedure can also be implemented using a stack, where each entry in the stack is
also a bit. First, in order to completely reconstruct the input sequence $X$, the decoder needs to know the statistics of the input sequence, including the total number $N$ of symbols, the frequency $f_X(s)$ of each symbol $s$ and the selected fixed interval $I$. Let $\tilde{c}_X(i)\triangleq s$ if $c_X(s)\leq i<c_X(s+1)$ for $i\in [0,N)$ and $s\in \Sigma_B$. Given a start state $x$ for decoding (i.e., the final state in the encoding) and the compressed bits stored in the stack, we pop $n$-bit digit $d$ from the stack and decode it to a symbol $s$, i.e., $s\gets \tilde{c}_X(d)$. Then, we update the state by $x\gets f_X(s)\times x+r$, where $r=d-c_X(s)$. It can be seen that after a symbol is decoded, the value of state becomes larger. As more and more symbols get decoded, its value will eventually grow beyond a bound. When the state $x$ exceeds the upper bound $2^T$ chosen during the encoding, i.e., $x\geq 2^T$, we push the least significant $v\times n$ bits of $x$ to the stack to make $x$ smaller, i.e., $x\gets x\gg v\times n$. We repeat
the above steps, and the repetition stops when the reconstruction of input sequence $X$ is achieved. Figure \ref{fig.3} shows a worked-through example for a partial sequence ``baabc" with parameters $n=8$, $v=2$ and $T=24$. The start state for encoding is chosen as $x=2^{T}-1=16777215$. It can be seen that the encoder and decoder go through the same state just in opposite order. Also, the order of the symbols reconstructed by the decoder is opposite to that received by the encoder, e.g., the last encoded symbol is decoded first.

\section{Proposed Coding Scheme}\label{sec:3}
The first limitation in our previous work~\cite{6194314} is that it uses arithmetic coding as the underlying coding framework. Arithmetic coding assigns each symbol to a sub-interval $[\overline{L},\overline{H})\subseteq [0.0,1.0)$ according to the probability of that symbol. When a symbol is encoded, the sub-interval narrows, and $\overline{L}$ and $\overline{H}$
move closer together. Because the machine system cannot represent infinite precision, as more and more symbols are encoded, $\overline{L}$ gradually approaches $\overline{H}$ and the two values eventually may coincide, then further data embedding is impossible. In this section, we first propose a coding method that employs our recent ANS variant \cite{9810728} to embed a message $W$ into the host sequence $X$. Then, the data extraction of the host sequence and the embedded message is given. Notably, this section considers the case where the statistics of the input sequence are known to the encoder and decoder, and refers to them as static data embedding and static data extraction. In the following, let $x_i$ and $y_i$ represent the values of host signal and stego signal at position $i$ in the sequence, respectively.

\subsection{Static Data Embedding}\label{sec:3.1}
Our recent ANS-Variant coding (see Section \ref{sec:2.2.2} for details) can provide the compression ratio close to that of arithmetic coding. More importantly, it does not suffer from the computing precision problem that may occur in arithmetic coding. As a result, we adopt ANS-Variant as the coding framework to embed a message $W$ into the host sequence $X=\left(x_1, x_2, \cdots, x_N\right)$. The framework consists of a combination of an encoder and a decoder of ANS-Variant coding, where the encoder is used to embed the data, and the decoder returns the corresponding stego signals.
From the description of ANS-Variant coding, one can see that in order to encode a host signal $x_i$ into the current state $x$, the ANS-Variant encoder outputs an $n$-bit value $c_{x_i}+(x\bmod {f_{x_i}})$, and if the state $x$ is less than the lower bound of the selected fixed interval, the encoder reads bits to increase it. This observation can be exploited to realize data embedding, i.e., one can choose to read bits from the message $W$ to increase the value of the state if necessary. In addition, the BFI algorithm (described in Section \ref{sec:2.2.1}) can estimate the optimal pmf of the stego sequence for a given host sequence. Therefore, we can apply the ANS-Variant decoder on the statistics of the stego sequence according to the $n$-bit value $c_{x_i}+(x\bmod {f_{x_i}})$ to decode the corresponding stego signal $y_i$. That is, the $n$-bit value $c_{x_i}+(x\bmod {f_{x_i}})$ produced by the ANS-Variant encoder serves as the temporary information to decode the corresponding stego signal $y_i$.

\begin{algorithm}[t]
	\caption{\label{alg.1} Proposed static embedding algorithm}
	\LinesNumbered
	\KwIn{A host sequence $X=\left(x_1, x_2, \cdots, x_N\right)$, the message $W$ to embed, the alphabet size $B$, the cumulative pmf $P_{CX}$.}
	\KwOut{A stego sequence $Y=\left(y_1, y_2, \cdots, y_N\right)$, the final state $x$.}
	Get the host statistics $F_X=\left\{f_X(s)| s\in \Sigma_B\right\}$ and $C_X=\left\{c_X(s)|s\in \Sigma_B\right\}$ from the $P_{CX}$\;
	Use the BFI algorithm to estimate the optimal $P_{CY}$\;
	Get the stego statistics $F_Y=\left\{f_Y(y)| y\in \Sigma_B\right\}$ and $C_Y=\left\{c_Y(y)|y\in \Sigma_B\right\}$ from the $P_{CY}$\;
	Set a start state $x\gets 1$\;
	\For{$i=1$ to $N$}{
		\If{$x< f_X(x_i)\times 2^{T-v\times n}$}{
			$D \gets Read_W(v\times n)$\;
			$x \gets (x\ll v\times n)+D$\;
		}
		$\eta_X=c_X(x_i)+(x\bmod {f_X(x_i)})$\;
		$x\gets \lfloor x/f_X(x_i) \rfloor$\;
		$y_i\gets \tilde{c}_Y(\eta_X)$\;
		$x\gets f_Y(y_i)\times x+\eta_X-c_Y(y_i)$\;
		\If{$x\geq 2^T$}{
			call $Write_W(x,v\times n)$\;
			$x \gets x\gg v\times n$\;
		}
	}
\end{algorithm}

The details of the embedding procedure are as follows. First, we require knowledge of the host statistics $F_X=\left\{f_X(s)| s\in \Sigma_B\right\}$ and $C_X=\left\{c_X(s)|s\in \Sigma_B\right\}$, and get the cumulative pmf $P_{CX}=\left\{P_{CX}(x)|x\in \Sigma_B \right\}$, where $P_{CX}(x)=\sum_{i=0}^{x}P_X(i)$ and $P_X(i)=\frac{f_X(i)}{N}$. Then, we apply the BFI algorithm on the $P_{CX}$ to estimate the optimal cumulative pmf $P_{CY}$ of the stego sequence $Y$. Once the distribution $P_{CY}=\left\{P_{CY}(y)|y\in \Sigma_B \right\}$ is known, we can further get the stego statistics $F_Y=\left\{f_Y(y)| y\in \Sigma_B\right\}$ and $C_Y=\left\{c_Y(y)|y\in \Sigma_B\right\}$, where $c_Y(y)=P_{CY}(y)\times N$ and $f_Y(y)=c_Y(y+1)-c_Y(y)$.
During the embedding, we process the host signals forward, i.e., from the first to the last signal. Given a start state $x$, for each host signal $x_i$, we first use an ANS-Variant encoder to encode it. Let $I:=\left [2^{T-v\times n},2^T \right )$ and $I_{x_i}:=\left [ f_X(x_i)\times 2^{T-v\times n},2^T \right )$. During the encoding, we have the following two cases. In the first, if the current state $x<f_X(x_i)\times 2^{T-v\times n}$, we read $v\times n$ bits from the message $W$ to increase $x$. That is, if the decimal representation of the $v\times n$ bits read is $D$, we have $x\gets (x\ll v\times n)+D$. After that, we maintain an $n$-bit temporal information $\eta_X=c_X(x_i)+(x\bmod {f_X(x_i)})$  and update the state $x$ by $x\gets \mathcal{C}\left(x,x_i\right)=\lfloor x/f_X(x_i) \rfloor$. It is noted that the $n$-bit temporary information produced by the host signal $x_i$ will be used in the following steps to decode the stego signal $y_i$ at the corresponding position. In the second, if the current state $x$ is within $I_{x_i}$, we maintain an $n$-bit temporal information $\eta_X=c_X(x_i)+(x\bmod {f_X(x_i)})$ and produce a new state $x$ by $x\gets \mathcal{C}\left(x,x_i\right)$. 
Next, we utilize the temporary information $\eta_X$ maintained in the previous step and the stego statistics $F_Y=\left\{f_Y(y)| y\in \Sigma_B\right\}$ and $C_Y=\left\{c_Y(y)|y\in \Sigma_B\right\}$ to decode the corresponding stego signal $y_i$.
Precisely, we transmit the temporary information $\eta_X$ to an ANS-Variant decoder to decode the stego signal $y_i$, i.e., $y_i\gets \tilde{c}_Y(\eta_X)$. Then, we update the state $x$ by $x\gets f_Y(y_i)\times x+r_{y_i}$, where $r_{y_i}=\eta_X-c_Y(y_i)$. From the description of ANS-Variant decoding in Section \ref{sec:2.2.2}, one can see that the updated state $x$ has the following two cases. In the first, If the value of state $x$ exceeds the upper bound $2^T$ chosen during the  encoding, we write the least significant $v\times n$ bits of $x$ to the message $W$ to make $x$ smaller, i.e., $x\gets x\gg v\times n$. In the second, if the updated state is exactly within the interval $I$, no further changes to the state are needed.

We repeat the above steps for the next host signal $x_{i+1}$ until there is no next host signal. As can be seen, an embedding cycle consists of a combination of an encoder and a decoder of ANS-Variant coding. Among them, the encoder aims to embed the data, and the decoder is designed to obtain the corresponding stego signal. Algorithm \ref{alg.1} presents the details. In Algorithm \ref{alg.1}, Line $1$ aims to get the host statistics. Lines $2$--$3$ first apply the BFI algorithm to estimate the optimal cumulative pmf $P_{CY}$ and then get the stego statistics from the $P_{CY}$. In Line $4$, the start state $x$ is chosen to be small enough, e.g., set to $1$, in order to make the encoder read the bits from the message $W$ as early as possible, thus increasing the amount of embedded data. Lines $6$--$10$ perform the encoding of ANS-Variant coding. In particular, Lines $7$--$8$ read bits from the message $W$ to enlarge the state if the current state is too small. Lines $11$--$15$ perform the decoding of ANS-Variant coding to generate the corresponding stego signal $y_i$. Finally, we get a stego sequence $Y=\left(y_1, y_2, \cdots, y_N\right)$ and a final state $x$. The final state $x$ needs to be transmitted along with the stego sequence $Y$, so that the extractor knows what state to start with. In addition, the shortened length (in bits) of the message $W$ is the amount of data embedded.

\subsection{Static Data Extraction}\label{sec:3.2}
In RDH techniques, data extraction includes extracting the host sequence and the embedded message from the received stego sequence. In the proposed static data embedding (see Section \ref{sec:3.1} for details), we first perform ANS-Variant encoding based on the known host statistics, and then perform ANS-Variant decoding under the premise that the optimal cumulative pmf of the stego sequence has been estimated. The data extraction is just the inverse of data embedding. Therefore, in order to achieve lossless data extraction, we first perform the ANS-Variant encoding under the premise that the optimal cumulative pmf of the stego sequence is known. Then, we perform the ANS-Variant decoding based on the host statistics to decode the host signal. It is worth noting that the data extraction requires transmitting the cumulative pmfs $P_{CX}$ and $P_{CY}$ as parameters. Another strategy is to only transmit $P_{CX}$ and the extractor generates the $P_{CY}$ with the BFI algorithm. The latter saves space but requires a computational cost to run the BFI algorithm. This section takes the latter as an example to introduce the process of static data extraction.

Given a stego sequence $Y=\left(y_1, y_2, \cdots, y_N\right)$, a start state $x$, and the cumulative pmf $P_{CX}=\left\{P_{CX}(x)|x\in \Sigma_B \right\}$, the details of data extraction are as follows. First, we apply the BFI algorithm on the host $P_{CX}$ to estimate the optimal cumulative pmf $P_{CY}=\left\{P_{CY}(y)|y\in \Sigma_B \right\}$. According to the $P_{CX}$ and $P_{CY}$, we calculate the stego statistics $F_Y=\left\{f_Y(y)| y\in \Sigma_B\right\}$ and $C_Y=\left\{c_Y(y)|y\in \Sigma_B\right\}$, and the host statistics $F_X=\left\{f_X(s)| s\in \Sigma_B\right\}$ and $C_X=\left\{c_X(s)|s\in \Sigma_B\right\}$, where $c_Y(y)=P_{CY}(y)\times N$,  $f_Y(y)=c_Y(y+1)-c_Y(y)$, and $c_X(s)=P_{CX}(s)\times N$,  $f_X(s)=c_X(s+1)-c_X(s)$. These statistics are used in the subsequent encoding and decoding of ANS-Variant coding. As the encoder and decoder of ANS-Variant coding run in reverse order, we process the stego signals backwards, i.e., from the last to the first signal. For each stego signal $y_i$, we first perform the ANS-Variant encoding based on the obtained stego statistics $F_Y$ and $C_Y$. Specifically, let $I:=\left [2^{T-v\times n},2^T \right )$ and $I_{y_i}:=\left [ f_Y(y_i)\times 2^{T-v\times n},2^T \right )$. In the encoding, we have the following two cases. In the first, if the current state $x<f_Y(y_i)\times 2^{T-v\times n}$, we read $v\times n$ bits from the message $W$ to increase $x$. Then, we hold an $n$-bit temporal information $\eta_Y=c_Y(y_i)+(x\bmod {f_Y(y_i)})$ and update the state $x$ by $x\gets \mathcal{C}\left(x,y_i\right)=\lfloor x/f_Y(y_i) \rfloor$.
In the second, if the current state $x$ is within $I_{y_i}$, we directly maintain an $n$-bit temporal information $\eta_Y=c_Y(y_i)+(x\bmod {f_Y(y_i)})$ and produce a new state $x$ by $x\gets \mathcal{C}\left(x,y_i\right)$. The above $n$-bit temporary information $\eta_Y$ produced by encoding the stego signal $y_i$ is used in the following steps to decode the corresponding host signal $x_i$.

\begin{algorithm}[t]
	\caption{\label{alg.2} Proposed static extraction algorithm}
	\LinesNumbered
	\KwIn{A stego sequence $Y=\left(y_1, y_2, \cdots, y_N\right)$, a start state $x$, the alphabet size $B$, the cumulative pmf $P_{CX}$.}
	\KwOut{A host sequence $X=\left(x_1, x_2, \cdots, x_N\right)$, the embedded message $W$.}
	Use the BFI algorithm to estimate the optimal $P_{CY}$\;
	Get the host statistics $F_X=\left\{f_X(s)| s\in \Sigma_B\right\}$ and $C_X=\left\{c_X(s)|s\in \Sigma_B\right\}$ from the $P_{CX}$\;
	Get the stego statistics $F_Y=\left\{f_Y(y)| y\in \Sigma_B\right\}$ and $C_Y=\left\{c_Y(y)|y\in \Sigma_B\right\}$ from the $P_{CY}$\;
	\For{$i=N$ to $1$}{
		\If{$x< f_Y(y_i)\times 2^{T-v\times n}$}{
			$D \gets Read_W(v\times n)$\;
			$x \gets (x\ll v\times n)+D$\;
		}
		$\eta_Y=c_Y(y_i)+(x\bmod {f_Y(y_i)})$\;
		$x\gets \lfloor x/f_Y(y_i) \rfloor$\;
		$x_i\gets \tilde{c}_X(\eta_Y)$\;
		$x\gets f_X(x_i)\times x+\eta_Y-c_X(x_i)$\;
		\If{$x\geq 2^T$}{
			call $Write_W(x,v\times n)$\;
			$x \gets x\gg v\times n$\;
		}
	}
\end{algorithm}

Next, we utilize the temporary information $\eta_Y$ maintained in the previous step and the host statistics $F_X=\left\{f_X(s)| s\in \Sigma_B\right\}$ and $C_X=\left\{c_X(s)|s\in \Sigma_B\right\}$ to decode the host signal $x_i$.
Precisely, we transmit the temporary information $\eta_Y$ to the ANS-Variant decoder to decode the host signal $x_i$, i.e., $x_i\gets \tilde{c}_X(\eta_Y)$. Then, we update the state $x$ by $x\gets f_X(x_i)\times x+r_{x_i}$, where $r_{x_i}=\eta_Y-c_X(x_i)$. From the description of ANS-Variant decoding in Section \ref{sec:2.2.2}, one can see that the updated state $x$ has the following two cases. In the first, if the value of state $x$ exceeds the upper bound $2^T$ chosen during the  data embedding, we write the least significant $v\times n$ bits of $x$ to the message $W$ to make $x$ smaller, i.e., $x\gets x\gg v\times n$. In the second, if the updated state is exactly within the interval $I$, no further changes to the state are needed. We repeat the above steps for the next stego signal $y_{i-1}$ until there is no next stego signal. 
Algorithm \ref{alg.2} gives the details. In Algorithm \ref{alg.2}, Line $1$ uses the BFI algorithm to estimate the optimal cumulative pmf $P_{CY}$.
Lines $2$--$3$ derive the host and stego statistics from the known $P_{CX}$ and $P_{CY}$, respectively. Lines $5$--$9$ perform the encoding of ANS-Variant coding on the stego signal and maintain a temporary information $\eta_Y$. Lines $10$--$14$ utilize the temporary $\eta_Y$ produced in the previous step to decode the host signal $x_i$. In particular, Lines $12$--$14$ write $v\times n$ bits to the message $W$ when the state exceeds a given upper bound $2^T$. Finally, we can reconstruct the host sequence $X=\left(x_1, x_2, \cdots, x_N\right)$ and the embedded message $W$ without loss.

\section{Dynamic Implementation}\label{sec:4}
Section \ref{sec:3} presents the embedding/extraction procedure to embed a message in the i.i.d. host sequence. One can see that the proposed static RDH scheme, like our previous work~\cite{6194314}, requires the cumulative pmf $P_{CX}$ of the host sequence as a parameter in both the embedding/extraction process.
Namely, the proposal involves using preset probabilities of signals that remain static during the whole process. However, transmitting the host pmf costs an additional overhead proportional to the alphabet size $B$, which is usually unacceptable in extremely space-constrained scenarios. Moreover, the usability of the proposal is based on the premise that the pmf of the host sequence is known, imposing certain limitations on the practical application.
In this section, we present a dynamic implementation without any prior knowledge of the host pmf. The implementation changes the probability after each host signal is encountered, thus saving the prior transmission of $P_{CX}$. First, we describe the dynamic data embedding algorithm. Second, the corresponding dynamic data extraction algorithm is given.

\subsection{Dynamic Data Embedding}\label{sec:4.1}
When any prior knowledge of the host pmf is not available, the proposed dynamic embedding method becomes a two-pass procedure. First, considering that the ANS-Variant decoder reconstructs all symbols in the reverse order of those processed by its encoder, so in order to ensure the synchronization of the encoder and the decoder, it is necessary to follow the order of the decoder run to collect the statistics of the host signal at each location. Therefore, we first pass the host sequence $X=\left(x_1, x_2, \cdots, x_N\right)$ once in the decoder's running order. Precisely, we start with a frequency distribution $F_X$, e.g., $F_X=\left\{f_X(s)=1, s\in \Sigma_B\right\}$, where $f_X(s)$ denotes the number of occurrence of the signal $s$. If the ANS-Variant decoder runs backwards, i.e., from the last to the first signal, we update the frequency distribution $F_X$ backwards based on the host signals encountered. Details of the update are as follows. We start with the last host signal $x_N$ and perform a backward pass over the host sequence. During the backward pass, when a host signal $x_i$ is encountered, the frequency corresponding to signal $x_i$ is updated via $f_X(x_i)\gets f_X(x_i)+1$. Next, we backward continue with the next signal and update its frequency in the same manner. When the frequency of the first host signal $x_1$ is updated and there is no next host signal, we get the newest frequency distribution $F_{X}$, which counts the frequency of each signal in the host sequence $X$ over the start frequency distribution.

\begin{algorithm}[t]
	\caption{\label{alg.3} Proposed dynamic embedding algorithm}
	\LinesNumbered
	\KwIn{A host sequence $X=\left(x_1, x_2, \cdots, x_N\right)$, a message $W$ to embed, the alphabet size $B$.}
	\KwOut{A stego sequence $Y=\left(y_1, y_2, \cdots, y_N\right)$, the final state $x$.}
	Initialize $F_X=\left\{f_X(s)=1| s\in \Sigma_B\right\}$\;
	\For{$i=N$ to $1$}{
		$f_X(x_i)\gets f_X(x_i)+1$\;
}
	Set a start state $x\gets 1$\;
	\For{$i=1$ to $N$}{
		$f_X(x_i)\gets f_X(x_i)-1$\;
		Get the host statistic $C_X=\left\{c_X(s)|s\in \Sigma_B\right\}$ and the cumulative pmf $P_{CX}$ from the newest $F_{X}=\left\{f_X(s)| s\in \Sigma_B\right\}$\;
		Use the BFI algorithm to estimate the optimal cumulative pmf $P_{CY}$\;
		Get the stego statistics $F_Y=\left\{f_Y(y)| y\in \Sigma_B\right\}$ and $C_Y=\left\{c_Y(y)|y\in \Sigma_B\right\}$ from the $P_{CY}$\;
		\If{$x< f_X(x_i)\times 2^{T-v\times n}$}{
			$D \gets Read_W(v\times n)$\;
			$x \gets (x\ll v\times n)+D$\;
		}
		$\eta_X=c_X(x_i)+(x\bmod {f_X(x_i)})$\;
		$x\gets \lfloor x/f_X(x_i) \rfloor$\;
		$y_i\gets \tilde{c}_Y(\eta_X)$\;
		$x\gets f_Y(y_i)\times x+\eta_X-c_Y(y_i)$\;
		\If{$x\geq 2^T$}{
			call $Write_W(x,v\times n)$\;
			$x \gets x\gg v\times n$\;
		}
	}
\end{algorithm}

When the first backward pass is completed, with the help of the newest statistics $F_{X}=\left\{f_{X}(s)| s\in \Sigma_B\right\}$, we then use an ANS-Variant encoder followed by an ANS-Variant decoder to perform data embedding. The dynamic embedding procedure is similar to that of the static data embedding described in Section \ref{sec:3.1}, except that the dynamic version employs a changing pmf. Given a host sequence $X=\left(x_1, x_2, \cdots, x_N\right)$, a message $W$ to embed and the newest frequency distribution $F_{X}=\left\{f_{X}(s)| s\in \Sigma_B\right\}$ obtained in the first backward pass, the embedding details are described below. Considering that the encoder and decoder of the ANS-Variant coding run in opposite directions, and the first pass in the same order as the decoder runs is backwards, we perform a second forward pass (i.e., from the first to the last signal) on the host sequence. For each host signal $x_i$ encountered, we first update the corresponding frequency of signal $x_i$ by $f_{X}(x_i)\gets f_{X}(x_i)-1$. Then, we compute the host statistics $C_{X}=\left\{c_{X}(s)=\sum_{i=-1}^{s-1}f_{X}(i)|s\in \Sigma_B\right\}$ and the cumulative pmf $P_{CX}=\left\{c_{X}(s)/c_{X}(B-1)|s\in \Sigma_B \right\}$ from the newest $F_{X}=\left\{f_{X}(s)| s\in \Sigma_B\right\}$. Next, we apply the BFI algorithm on the $P_{CX}$ to estimate the optimal cumulative pmf $P_{CY}$ of the stego sequence $Y$. Once the $P_{CY}=\left\{P_{CY}(y)|y\in \Sigma_B \right\}$ is known, we can also get the stego statistics $F_{Y}=\left\{f_{Y}(y)| y\in \Sigma_B\right\}$ and $C_{Y}=\left\{c_{Y}(y)|y\in \Sigma_B\right\}$, where $c_{Y}(y)=P_{CY}(y)\times N$ and $f_{Y}(y)=c_{Y}(y)-c_{Y}(y-1)$. Once the host and stego statistics (i.e., $F_{X}$, $C_{X}$, $F_{Y}$ and $C_{Y}$) are both available, we then implement data embedding using similar steps in static data embedding (see Lines $6$--$15$ in Algorithm \ref{alg.1} for details).
We repeat the above steps for the next host signal until there is no next host signal. Finally, we get a stego sequence $Y=\left(y_1, y_2, \cdots, y_N\right)$ and a final frequency distribution $F_{X}$. From the update rule of the $F_{X}$, 
it can be seen that the final $F_{X}$ is exactly the same as the start $F_{X}$ before the first backward pass. Algorithm \ref{alg.3} gives the details. In Algorithm \ref{alg.3}, Lines $1$--$3$ initialize the frequency distribution $F_{X}$ to all ones and perform a backward pass. Line $6$ updates the frequency corresponding to the signal $x_i$. Lines $7$--$8$ first calculate the host statistics $C_X$ and $P_{CX}$, and then use the BFI algorithm to estimate the optimal $P_{CY}$. Line $9$ gets the stego statistics $F_Y$ and $C_Y$ from the $P_{CY}$.
Lines $10$--$14$ perform the encoding of ANS-Variant coding. In particular, Lines $11$--$12$ read bits from the message $W$ to enlarge the state if it is too small. Lines $15$--$19$ perform the decoding of ANS-Variant coding to obtain the corresponding stego signal $y_i$. Finally, 
we get a stego sequence $Y=\left(y_1, y_2, \cdots, y_N\right)$ and a final state $x$.

According to the above description, it is clear that the host pmf is not necessary in the proposed dynamic data embedding. Therefore, unlike our previous work~\cite{6194314}, the proposed dynamic method overcomes the limitation of explicitly transmitting the host pmf $P_{CX}$, and brings flexibility in practical applications.

\subsection{Dynamic Data Extraction}\label{sec:4.2}
In order to reconstruct the host sequence and the embedded message from the stego sequence without any prior knowledge of the host pmf, the proposed dynamic data extraction also initializes a frequency distribution $F_{X'}$ and changes it after a host signal is decoded. The start $F_{X'}$ is exactly the same as the start $F_{X}$ before the first backward pass in the dynamic data embedding, i.e., we have $F_{X'}=\left\{f_{X'}(s)=1, s\in \Sigma_B\right\}$, where $f_{X'}(s)$ denotes the number of occurrence of signal $s$. The dynamic extraction procedure is similar to that of the static data extraction described in Section \ref{sec:3.2}, except that the dynamic version employs a changing pmf to save the prior transmission of the host pmf.

Given a stego sequence $Y=\left(y_1, y_2, \cdots, y_N\right)$, a final state $x$ in the dynamic embedding, the details of the dynamic data extraction are as follows. First, we initialize the frequency distribution $F_{X'}$ as $F_{X'}=\left\{f_{X'}(s)=1, s\in \Sigma_B\right\}$. Then, 
considering that the encoder and decoder of the ANS-Variant coding run in the opposite directions, we process the stego signals backwards, i.e., from the last to the first signal. For each stego signal $y_i$, we first get the statistics $C_{X}=\left\{c_{X}(s)|s\in \Sigma_B\right\}$ and the cumulative pmf $P_{CX}=\left\{P_{CX}(s)| s\in \Sigma_B\right\}$ from the newest $F_{X'}=\left\{f_{X'}(s)| s\in \Sigma_B\right\}$, where $c_{X}(s)= \sum_{i=-1}^{s-1}f_{X'}(i)$ and $P_{CX}(s)=c_{X}(s)/c_{X}(B-1)$. Then, we adopt the BFI algorithm on the $P_{CX}$ to estimate the optimal cumulative pmf $P_{CY}=\left\{P_{CY}(y)|y\in \Sigma_B \right\}$. Once the $P_{CY}$ is available, the corresponding stego statistics $F_Y=\left\{f_Y(y)| y\in \Sigma_B\right\}$ and $C_Y=\left\{c_Y(y)|y\in \Sigma_B\right\}$ can also be calculated, where $c_Y(y)= P_{CY}(y)\times N$ and $f_Y(y)=c_Y(y+1)-c_Y(y)$. Next, we implement data extraction using similar steps in static data extraction (see Lines $5$--$14$ in Algorithm \ref{alg.2} for details). Notably, when a host signal $x_i$ is decoded, we need to update its frequency by $f_{X'}(x_i)\gets f_{X'}(x_i)+1$. 
We repeat the above steps for the next stego signal $y_{i-1}$ until there is no next stego signal. Finally, we can reconstruct the host sequence and the embedded message.
Algorithm \ref{alg.4} gives the details. In Algorithm  \ref{alg.4}, Line $1$ initializes $F_{X'}$ to all ones. Line $3$ gets the host statistics $C_X$ and $P_{CX}$ from the newest $F_{X'}$. Lines $4$--$5$ first use the BFI algorithm to estimate the optimal $P_{CY}$ and then get the stego statistics $F_Y$ and $C_Y$ from the $P_{CY}$. Lines $6$--$10$ perform the ANS-Variant encoding on the stego signals and maintain a temporary information $\eta_Y$. Lines $11$--$12$ utilize the temporary $\eta_Y$ to decode the host signal $x_i$ and update the state $x$. If the updated state exceeds the selected upper bound $2^T$, Lines $14$--$15$ write the least significant $v\times n$ bits of $x$ to the message $W$ to decrease $x$. Finally, Line $16$ updates the frequency of the corresponding host signal $x_i$.
From the above description, one can see that the dynamic data extraction procedure does not require any prior knowledge of the host pmf. It dynamically adjusts the pmf based on the decoded host signals to save the additional overhead.

\begin{algorithm}[t]
	\caption{\label{alg.4} Proposed dynamic extraction algorithm}
	\LinesNumbered
	\KwIn{A stego sequence $Y=\left(y_1, y_2, \cdots, y_N\right)$, a final state $x$, the alphabet size $B$.}
	\KwOut{A host sequence $X=\left(x_1, x_2, \cdots, x_N\right)$, the embedded message $W$.}
	Initialize $F_{X'}=\left\{f_{X'}(s)=1| s\in \Sigma_B\right\}$\;
	\For{$i=N$ to $1$}{
		Get the host statistic $C_X=\left\{c_X(s)|s\in \Sigma_B\right\}$ and the cumulative pmf $P_{CX}$ from the $F_{X'}=\left\{f_{X'}(s)| s\in \Sigma_B\right\}$\;
		Use the BFI algorithm to estimate the optimal $P_{CY}$\;
		Get the stego statistics $F_Y=\left\{f_Y(y)| y\in \Sigma_B\right\}$ and $C_Y=\left\{c_Y(y)|y\in \Sigma_B\right\}$ from the $P_{CY}$\;
		\If{$x< f_Y(y_i)\times 2^{T-v\times n}$}{
	$D \gets Read_W(v\times n)$\;
	$x \gets (x\ll v\times n)+D$\;
}
	$\eta_Y=c_Y(y_i)+(x\bmod {f_Y(y_i)})$\;
	$x\gets \lfloor x/f_Y(y_i) \rfloor$\;
	$x_i\gets \tilde{c}_X(\eta_Y)$\;
	$x\gets f_X(x_i)\times x+\eta_Y-c_X(x_i)$\;
	\If{$x\geq 2^T$}{
		call $Write_W(x,v\times n)$\;
		$x \gets x\gg v\times n$\;
	}
   $f_{X'}(x_i)\gets f_{X'}(x_i)+1$\;
}
\end{algorithm}

\section{Simulations}\label{sec:5}
In this section, we evaluate the embedding performance of the proposed static/dynamic method in terms of the mean squared error (MSE), embedding rate (in bits per pixel, bpp) and
PSNR value (in dB) for images.
The proposed methods are compared with three methods, including our previous work~\cite{6194314} and two high capacity RDH methods \cite{8733828}, \cite{JIA2019238}. The proposed methods are implemented in Visual C++ code with OpenCV image development kits. First, we show the experimental results on the i.i.d. sequence. Second, all experiments are conducted on gray-scale images (two common standard images of size 512 $\times$ 512).

\begin{figure}[t]
	\centering
	\includegraphics[width=3.0in]{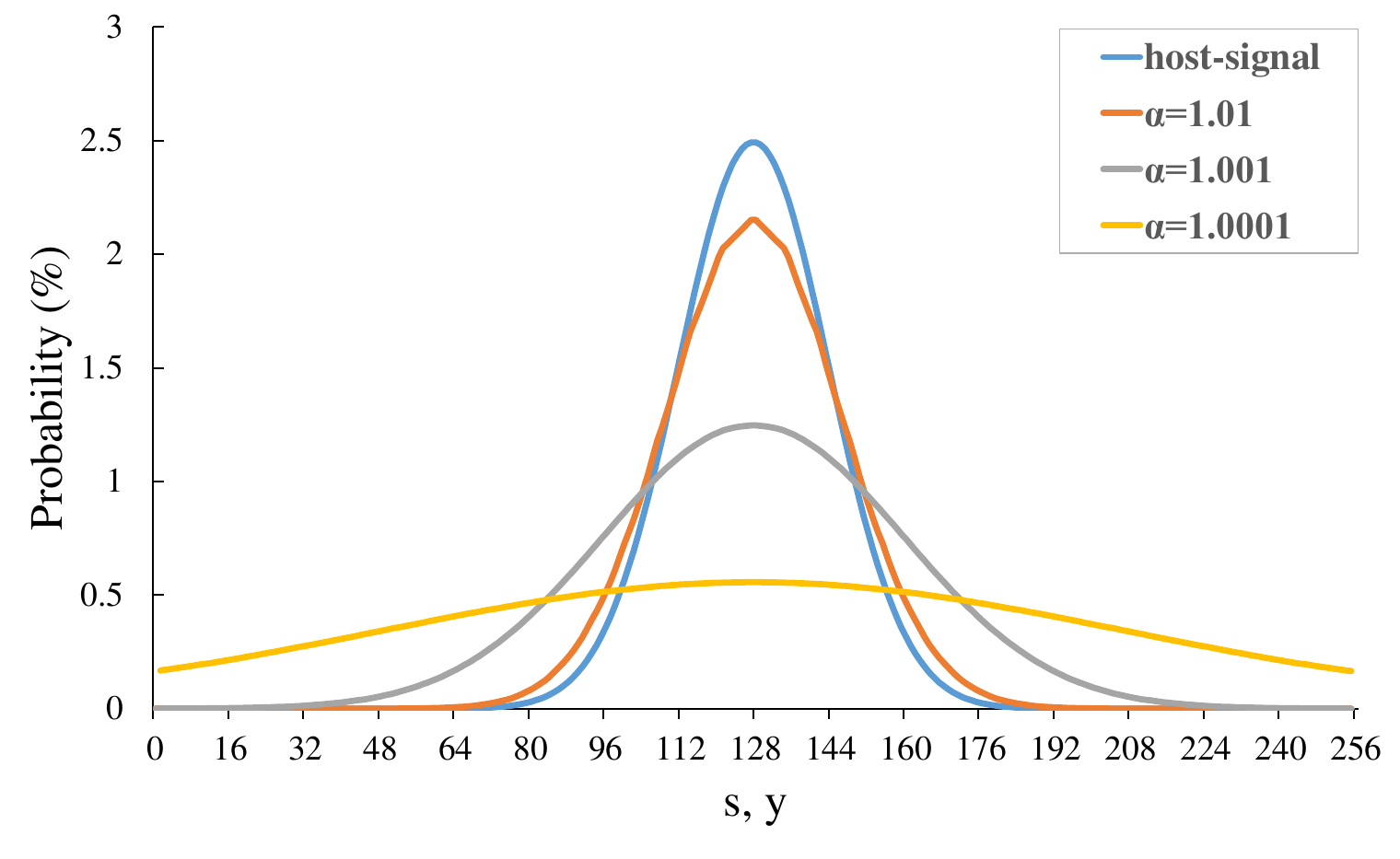}
	\caption{The pmf of host signals drawn from discrete normal distribution versus pmf of the corresponding stego-signals for various $\alpha$.}
	\label{fig.4}
\end{figure}

\subsection{Experiments for i.i.d. Sequence}\label{sec:5.1}
In this subsection, we test the embedding performance of the proposed static/dynamic method and our previous work~\cite{6194314} on i.i.d. sequence. We consider the host sequence drawn from discrete normal distribution. The host signals are $8$-bit gray-scale with $B=256$, and the mean of the normal distribution is at $127.5$. Figure \ref{fig.4} shows the pmf of host signals for the standard deviation $\sigma=256$, and the pmf of stego signals by applying the BFI algorithm for $\alpha=1.01, 1.001$ and $1.0001$. It can be seen that the pmf of stego signals are more stationary than that of host signals, thus the stego sequence has a larger Shannon entropy. That is, more information is required on average to represent it, and the increased volume can be utilized to embed secret data.

\begin{figure}[t]
	\centering
	\includegraphics[width=3.0in]{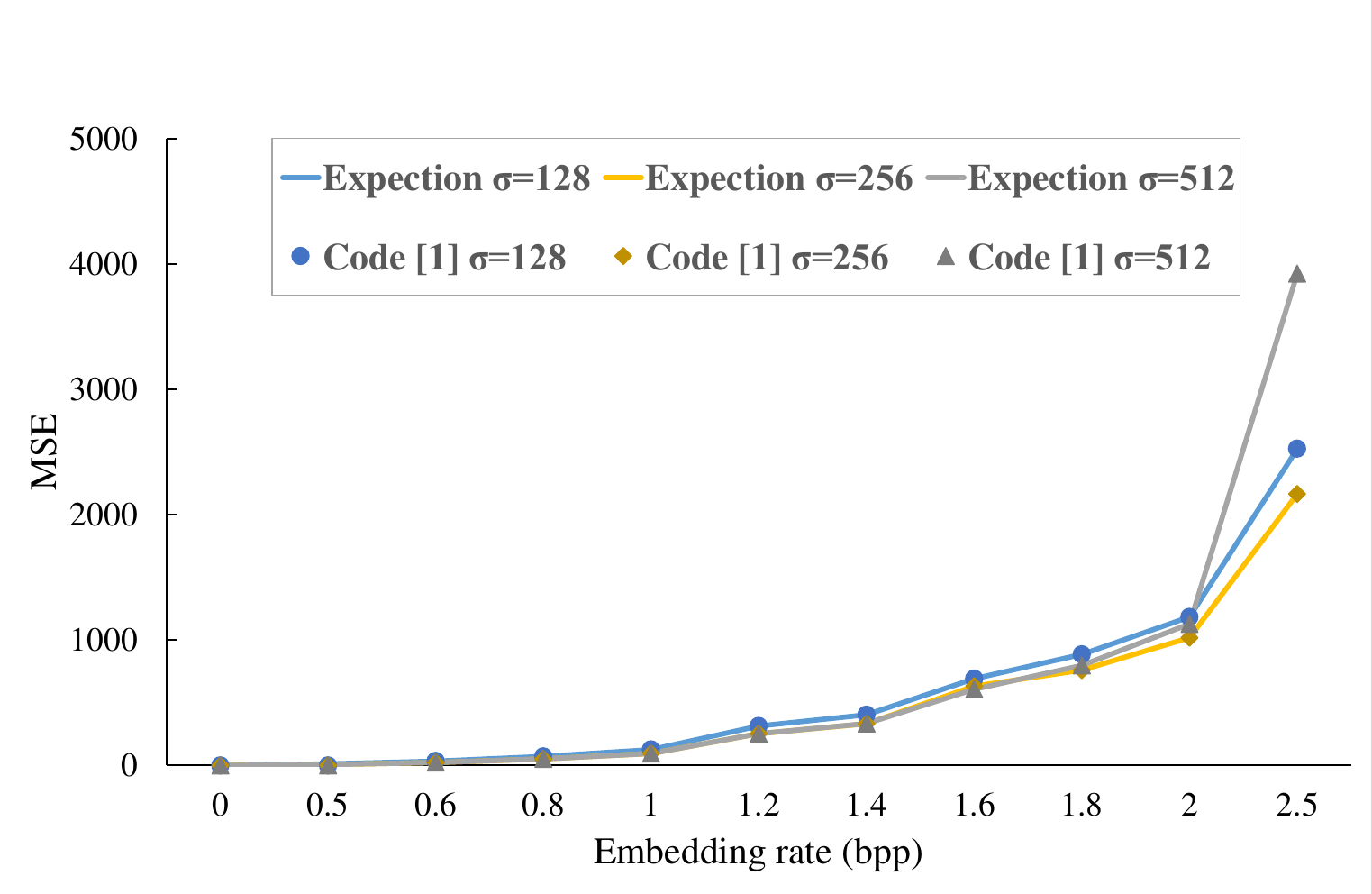}
	\caption{Rate-distortion curves of a given normal distribution for various $\sigma$, and rate-distortion values of our previous code \cite{6194314}.}
	\label{fig.6}
\end{figure}

\begin{figure}[t]
	\centering
	\includegraphics[width=3.0in]{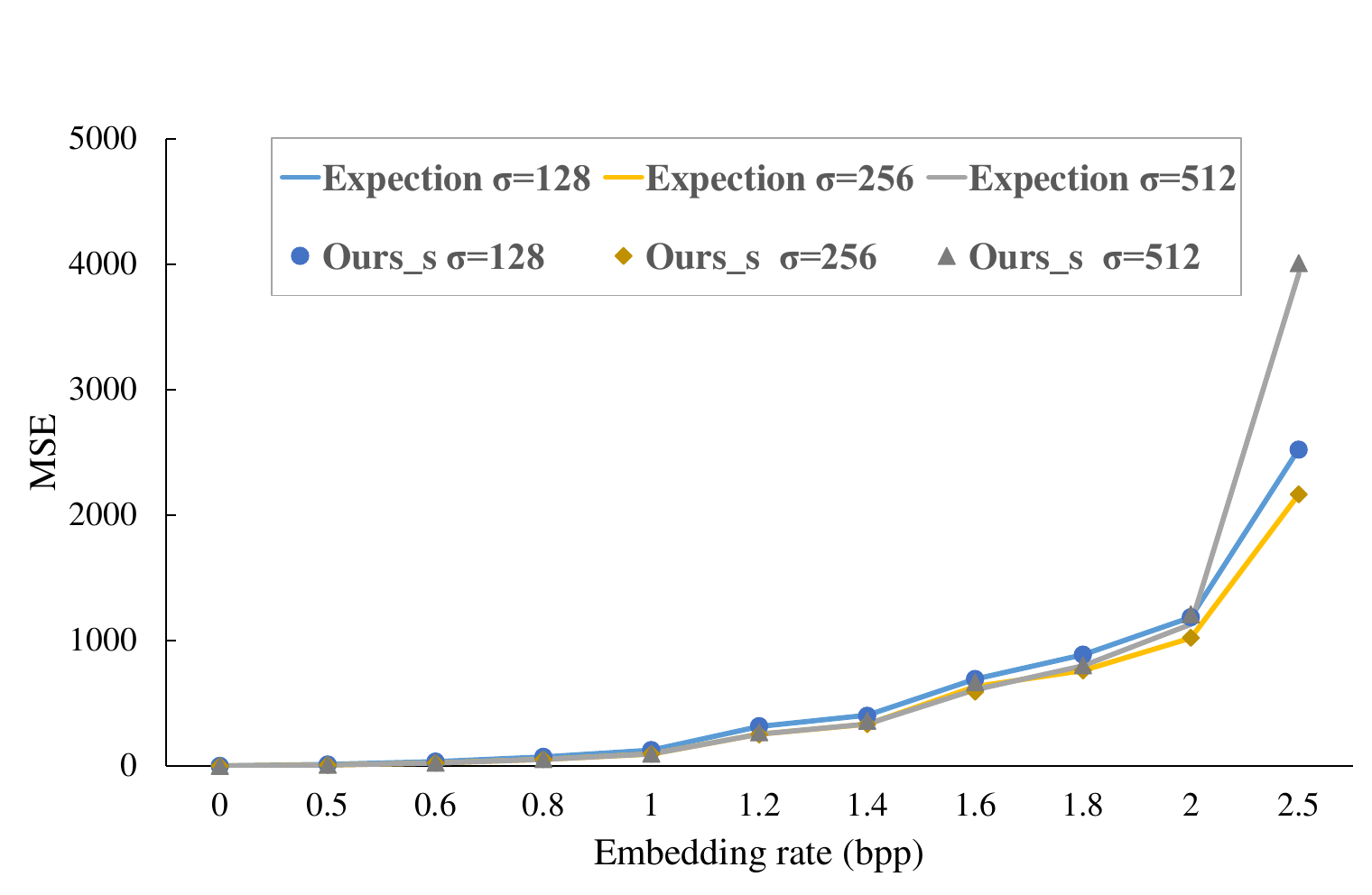}
	\caption{Rate-distortion curves of a given normal distribution for various $\sigma$, and rate-distortion values of the proposed static method.}
	\label{fig.7}
\end{figure}

\begin{figure}[t]
	\centering
	\includegraphics[width=3.0in]{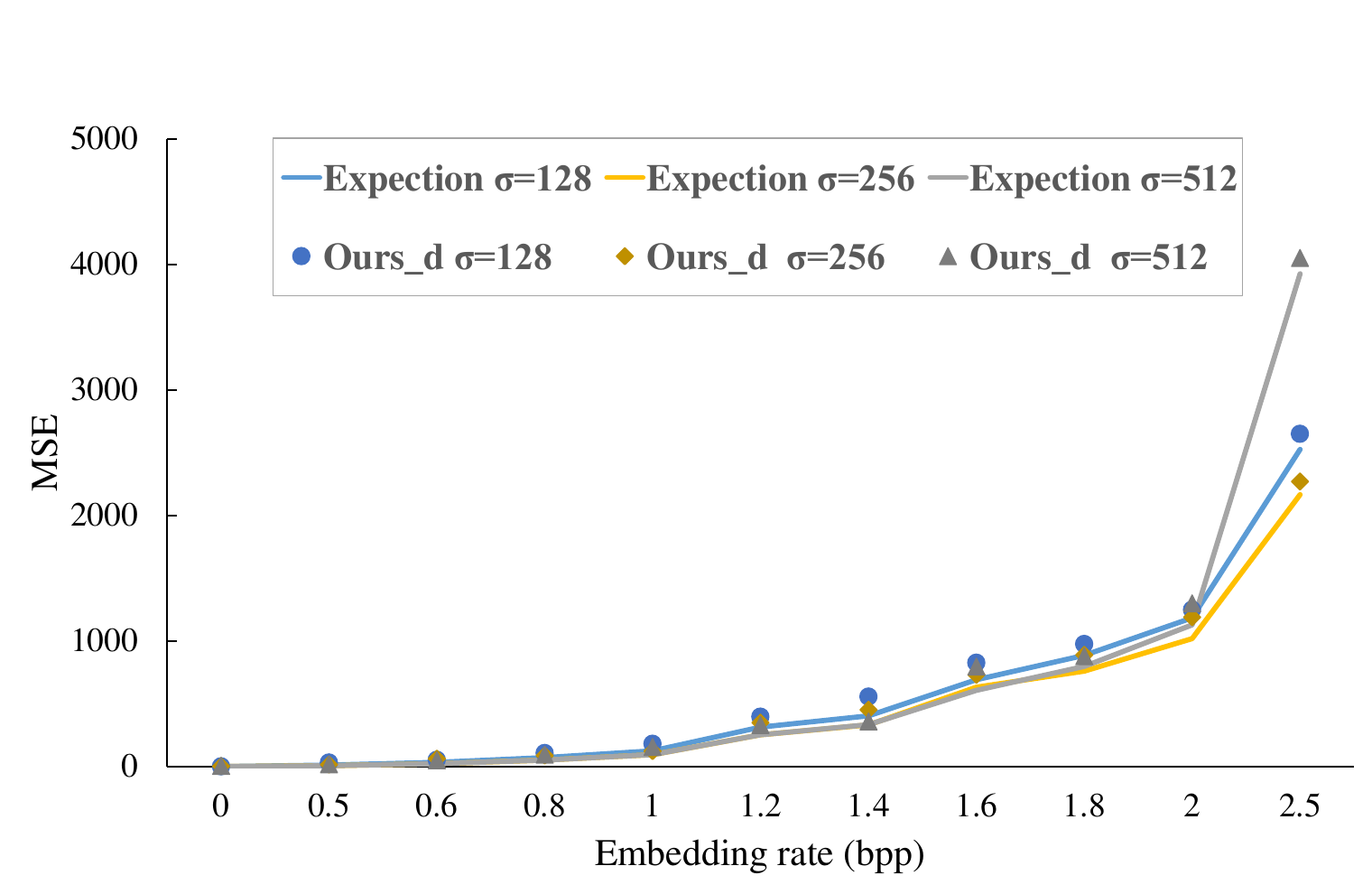}
	\caption{Rate-distortion curves of a given normal distribution for various $\sigma$, and rate-distortion values of the proposed dynamic method.}
	\label{fig.8}
\end{figure}

We adopt two common metrics embedding rate (bpp) and MSE to evaluate the embedding performance. For a host sequence $X=\left(x_1, x_2, \cdots, x_N\right)$ and a stego sequence $Y=\left(y_1, y_2, \cdots, y_N\right)$, the MSE is defined as
\[
\text{MSE}=\frac{1}{N}\sum_{i=1}^N(x_i-y_i)^2.
\]
The lower the MSE value, the better the quality.
The secret message $W$ is generated by multiple calls to the operation $RAND()*65536$, which returns a pseudo-random number between $0$ (inclusive) and $65536$ (exclusive). 
Figures \ref{fig.6}--\ref{fig.8} show the rate-distortion curves for the pmf  $P_{CX}$ with various $\sigma$. The blue line, yellow line and gray line, respectively, depict the expected rate-distortion curves for $\sigma=128$, $256$ and $512$. We implement our previous work \cite{6194314} (Code \cite{6194314}), the proposed static method (Ours\_s) and the proposed dynamic method (Ours\_d) to embed the message in the  $65536$ host signals for each corresponding pmfs $P_{CX}$ and $P_{CY}$, and the rate-distortion values are respectively marked as circles, rhombuses and triangles in Figures \ref{fig.6}--\ref{fig.8}. In the proposed static/dynamic implementation, we use $32$ bits to represent the state and choose parameters $T=16$, $n=16$ and $v=1$. From Figures \ref{fig.6}--\ref{fig.7}, one can see that the performance of the proposed static method is similar to that of our previous work \cite{6194314}, both approach the expected rate-distortion bound. More importantly, the proposed method uses ANS coding instead of arithmetic coding for RDH, which does not cause computing precision problems. Figure \ref{fig.8} gives the performance of the proposed dynamic method. It shows that the rate-distortion values in the dynamic version have a small distance from the expected rate-distortion bound. This is due to the fact that we use a predicted pmf instead of the true pmf. Especially, our proposal does not require transmitting the host pmf during data embedding and extraction, which saves additional overhead and enhances practicality.

\begin{figure}[t]
	\centering
	\subfloat[]{
		\includegraphics[width=4.0cm]{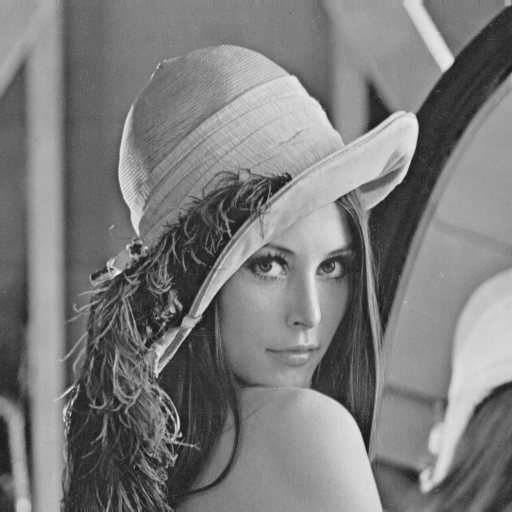}
		\label{hostImg}
	}
	\subfloat[]{
		\includegraphics[width=4.0cm]{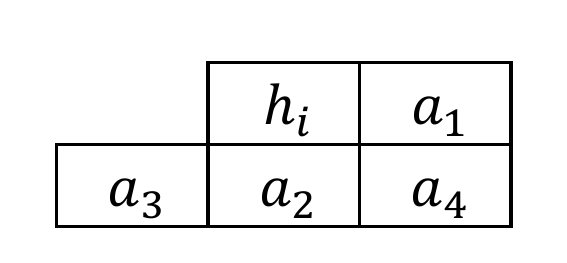}
		\label{predictor}
	}
	\caption{(a). Host image Lena. (b). Four neighboring pixels adopted in the predictor.}
	\label{image and predictor}
\end{figure}

\subsection{Experiments for Gray-Scale Images}\label{sec:5.2}
In this subsection, the embedding performance of the proposed static/dynamic method on gray-scale images is compared with three state-of-the-art methods, including our previous work~\cite{6194314} and two high capacity RDH methods~\cite{8733828} and \cite{JIA2019238}. To reduce the correlation of the neighboring image pixels, we preprocess the host image with predictive coding. The predictive coding processes the host pixels from left
to right, and from top to bottom. Figure \ref{image and predictor} shows a $512 \times 512$ host image Lena and depicts the four neighboring pixels used for predicting the host pixel $h_i$. The predicted value $\hat{h}_i$ is defined as
\begin{equation}\label{predict}
	\hat{h}_i= \frac{3}{8}\times a_1+ \frac{3}{8}\times a_2+\frac{1}{8}\times a_3+\frac{1}{8}\times a_4.
\end{equation}
For the pixels at the leftmost column, e.g, $a_3$, we pick the
nearby right pixel $a_2$ as the predicted value. For the pixels at the rightmost column, e.g, $a_1$, we pick the nearby lower pixel $a_4$ as the predicted value. For the pixels at bottom row, e.g, $a_2$, we pick the nearby right pixel $a_4$ as the predicted value. We omit the bottom-right corner pixel and default it to $128$.

In order to embed a message into the host image, the existing RDH methods usually utilize the differences between the host and predicted pixels. However, these methods use Modulo $256$ operation to avoid the overflow/underflow problem, i.e., the values of some pixels in the stego
image may exceed the upper or lower bound, e.g., $255$
or $0$ for an $8$-bit gray-scale image. This may cause the stego image to suffer from the salt and pepper noise. To overcome this issue, we suggest to develop a two-dimensional lookup table $\textit{K}$ of size $B\times B$. The row index $i$ of the table denotes the predicted pixel and the column index $j$ denotes the true host pixel, where $0\leq i<B$, $0\leq j<B$. The entry $\textit{K}[i][j]$ at the $i$-th row and the $j$-th column in the table $\textit{K}$ represents the number of host pixel $j$ predicted as predicted pixel $i$. Next, we give the details of static data embedding/extraction on gray-scale images. First, 
we initialize the two-dimensional table $\textit{K}$ to all ones, i.e., $\textit{K}[i][j]=1$, where $i, j\in \Sigma_B$. 
Then, for each host pixel $h_i$, we use the predictor to get its predicted value $\hat{h}_i$, and update the corresponding entry by $\textit{K}[\hat{h}_i][h_i] \gets \textit{K}[\hat{h}_i][h_i]+1$. If the predictor works accurately enough, i.e., $\hat{h}_i \approx h_i$, one can see that the larger values in the table $\textit{K}$ will be clustered on the diagonal. When the two-dimensional table $\textit{K}$ has been built, we then embed the message into host image by utilizing the corresponding row as the host pmf. We process the host pixels from left
to right, and from top to bottom. Specifically, for each host pixel $h_i$, we first use the predictor to get its predicted value $\hat{h}_i$. Then, we take $\hat{h}_i$ as the row index to look up the table $\textit{K}$, and return a one-dimensional statistic $\textit{K}[\hat{h}_i]$ of size $B$. Afterwards, the $\textit{K}[\hat{h}_i]$ is used as the frequency distribution to calculate the $P_{CX}$ in Algorithm \ref{alg.1}. Once the $P_{CX}$ is available, we can embed the message by applying Algorithm \ref{alg.1}. We repeat the above steps until there is no next host pixel. Finally, we can get a stego image. 
In the static data extraction, we reconstruct the host image in the reverse order from right to left, and from bottom to top. Consider the step of decoding $h_i$ shown in Figure \ref{predictor}, the four neighboring host pixels $a_1$, $a_2$, $a_3$ and $a_4$ had been decoded by the previous decoding steps. Therefore, we have the predicted value $\hat{h}_i$ through \eqref{predict}. Then, we take $\hat{h}_i$ as the row index to look up the table $\textit{K}$ and return a one-dimensional statistic $\textit{K}[\hat{h}_i]$ of size $B$. Also, the $\textit{K}[\hat{h}_i]$ is used as the frequency distribution to calculate the $P_{CX}$ in Algorithm \ref{alg.2}. Once the $P_{CX}$ is available, we can perform the data extraction via Algorithm \ref{alg.2}. We repeat the above steps until there is no next stego pixel. Finally, we can reconstruct the host image and embedded message.

Below, we discuss the details of dynamic data embedding/extraction on gray-scale images. First, we initialize a two-dimensional table $\textit{K}$ to all ones, i.e., $\textit{K}[i][j]=1$, where $i, j\in \Sigma_B$. Then, 
we perform a backward pass over the host image, and for each host pixel $h_i$, we use the predictor to obtain its predicted value $\hat{h}_i$. And, we update the corresponding entry in the table by $\textit{K}[\hat{h}_i][h_i] \gets \textit{K}[\hat{h}_i][h_i]+1$. We repeat the above steps until there is no next host pixel. Next, the dynamic data embedding is implemented on the basis of the newest table $\textit{K}$. We process the host pixels from left
to right, and from top to bottom. Specifically, for each host pixel $h_i$, we first use the predictor to get its predicted value $\hat{h}_i$. Notably, when a predicted value $\hat{h}_i$ is obtained, we update the corresponding entry by $\textit{K}[\hat{h}_i][h_i] \gets \textit{K}[\hat{h}_i][h_i]-1$. Then, we take $\hat{h}_i$ as the row index to look up the table $\textit{K}$ and return a one-dimensional statistic $\textit{K}[\hat{h}_i]$ of size $B$.
Afterwards, the $\textit{K}[\hat{h}_i]$ is used as the frequency distribution to calculate the host statistics $C_{X}$ and $P_{CX}$ in Line $7$ of Algorithm \ref{alg.3}. Then, we can embed the message via Lines $8$--$19$ in Algorithm \ref{alg.3}. We repeat the above steps until there is no next host pixel, and finally we can get a stego image.
In the corresponding dynamic data extraction, we first initialize a two-dimensional lookup table $\textit{K}$ to all ones, and predict the pixel $\hat{h}_i$ in the reverse order from right to left, and from bottom to top. Similarly, we then take $\hat{h}_i$ as the row index to look up the table $\textit{K}$ and return a statistic $\textit{K}[\hat{h}_i]$ of size $B$. The $\textit{K}[\hat{h}_i]$ is used to calculate the host statistics $C_{X}$ and $P_{CX}$ in Line $3$ of Algorithm \ref{alg.4}. Then, we can decode the host pixel $h_i$ via Lines $4$--$15$ in Algorithm \ref{alg.4}. 
Notably, when a host pixel $h_i$ is decoded, we update the corresponding entry by $\textit{K}[\hat{h}_i][h_i] \gets \textit{K}[\hat{h}_i][h_i]+1$. We repeat the above steps until all host pixels are decoded. Finally, the host image and the embedded message are reconstructed.
 
The PSNR (in dB) is often used as an objective
measure of image quality, which reflects the degree of similarity between the stego image and host image. That is, the larger the value of PSNR, the better the image quality.
The PSNR is calculated as
\[
\text{PSNR}=10\times \log 10(MAX^2/\text{MSE}).
\]
where $MAX$ is the maximum possible pixel value of the image (e.g., $255$ for $8$-bit gray-scale images), and MSE is the mean squared error between the host and stego image. Given a noise-free $L_1\times L_2$ host image $X$ and its stego image $Y$, MSE is defined as
\[
\text{MSE}=\frac{1}{L_1\times L_2}\sum_{i=0}^{L_1-1}\sum_{j=0}^{L_2-1}(X(i,j)-Y(i,j))^2,
\]
where $X(i,j)$ and $Y(i,j)$ represent the pixels at the 
$i$-th row and the $j$-th column in images $X$ and $Y$, respectively.

\begin{figure*}[htbp]
	\centering
	\subfloat[]{
		\begin{minipage}[t]{0.25\linewidth}
			\centering
			\includegraphics[width=1.5in]{lena.png}
		\end{minipage}%
	}%
	\subfloat[]{
		\begin{minipage}[t]{0.25\linewidth}
			\centering
			\includegraphics[width=1.5in]{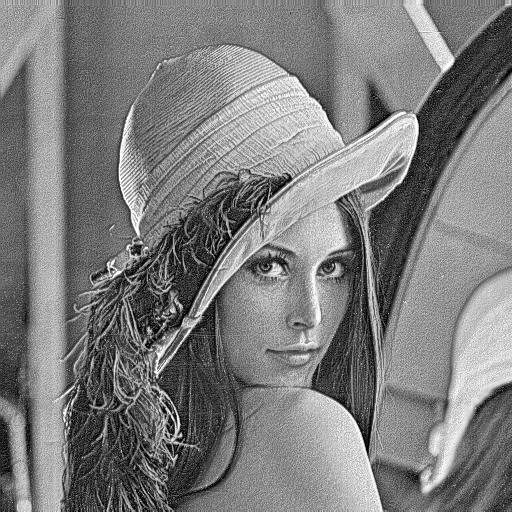}
		\end{minipage}%
	}%
	\subfloat[]{
		\begin{minipage}[t]{0.25\linewidth}
			\centering
			\includegraphics[width=1.5in]{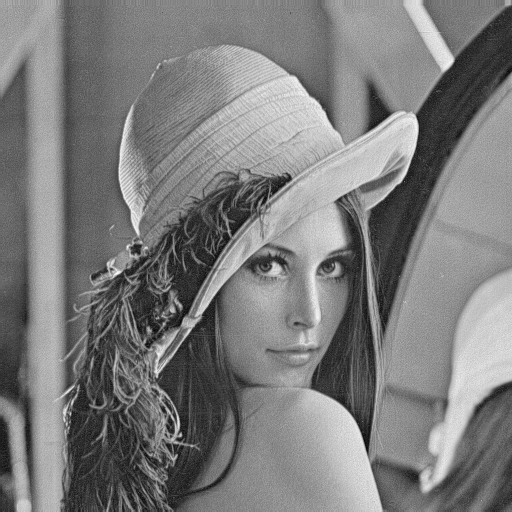}
		\end{minipage}
	}%
	\subfloat[]{
		\begin{minipage}[t]{0.25\linewidth}
			\centering
			\includegraphics[width=1.5in]{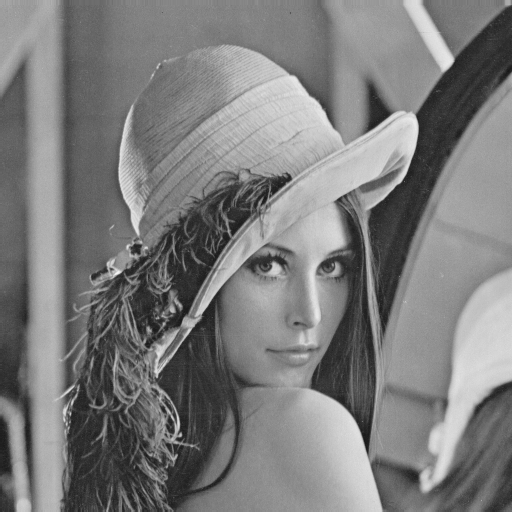}
		\end{minipage}
	}%
	\centering
	\caption{(a). Host image Lena. (b). Stego image, R=1.83, PSNR=24.2. (c). Stego image, R=1.07, PSNR=34.4. (d). Stego image, R=0.11, PSNR=56.5.}
	\label{lenatest}
\end{figure*}

\begin{figure*}[htbp]
	\centering
	\subfloat[]{
		\begin{minipage}[t]{0.25\linewidth}
			\centering
			\includegraphics[width=1.5in]{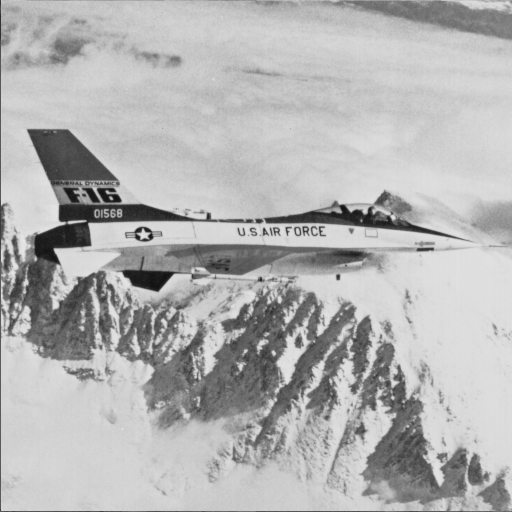}
		\end{minipage}%
	}%
	\subfloat[]{
		\begin{minipage}[t]{0.25\linewidth}
			\centering
			\includegraphics[width=1.5in]{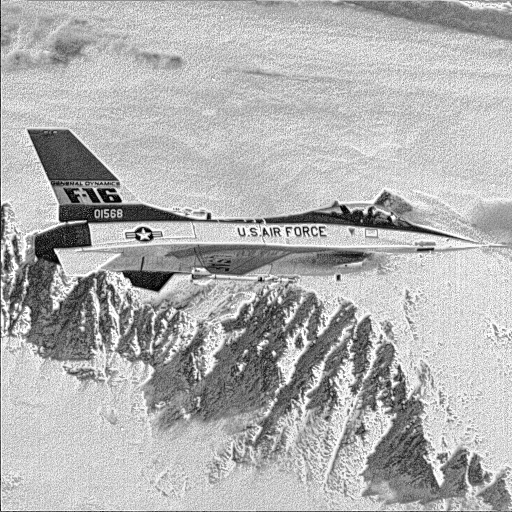}
		\end{minipage}%
	}%
	\subfloat[]{
		\begin{minipage}[t]{0.25\linewidth}
			\centering
			\includegraphics[width=1.5in]{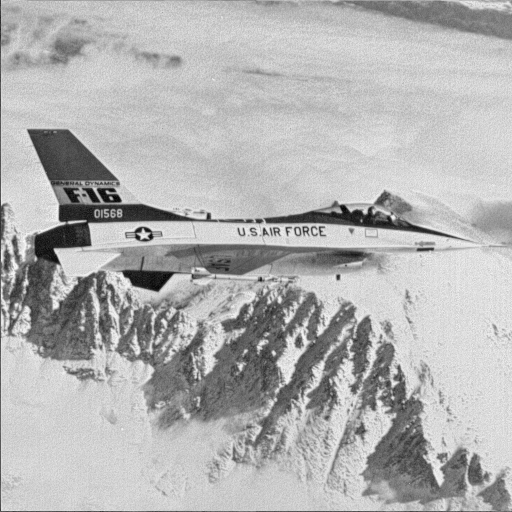}
		\end{minipage}
	}%
	\subfloat[]{
		\begin{minipage}[t]{0.25\linewidth}
			\centering
			\includegraphics[width=1.5in]{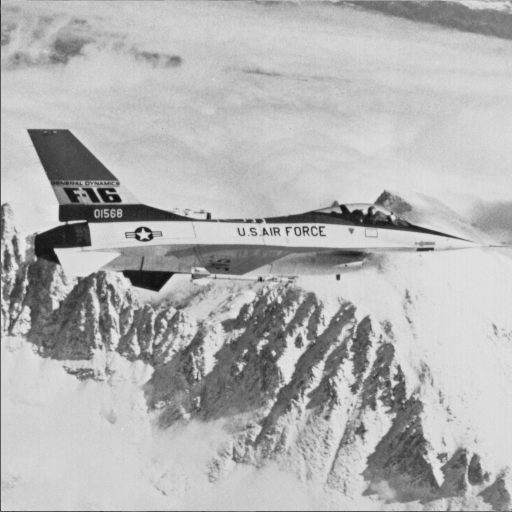}
		\end{minipage}
	}%
	\centering
	\caption{(a). Host image Airplane. (b). Stego image, R=2.65, PSNR=22.68. (c). Stego image, R= 1.46, PSNR=33.7. (d). Stego image, R=0.3, PSNR=53.0.}
	\label{planetest}
\end{figure*}

\begin{figure}[t]
	\centering
	\includegraphics[width=3.0in]{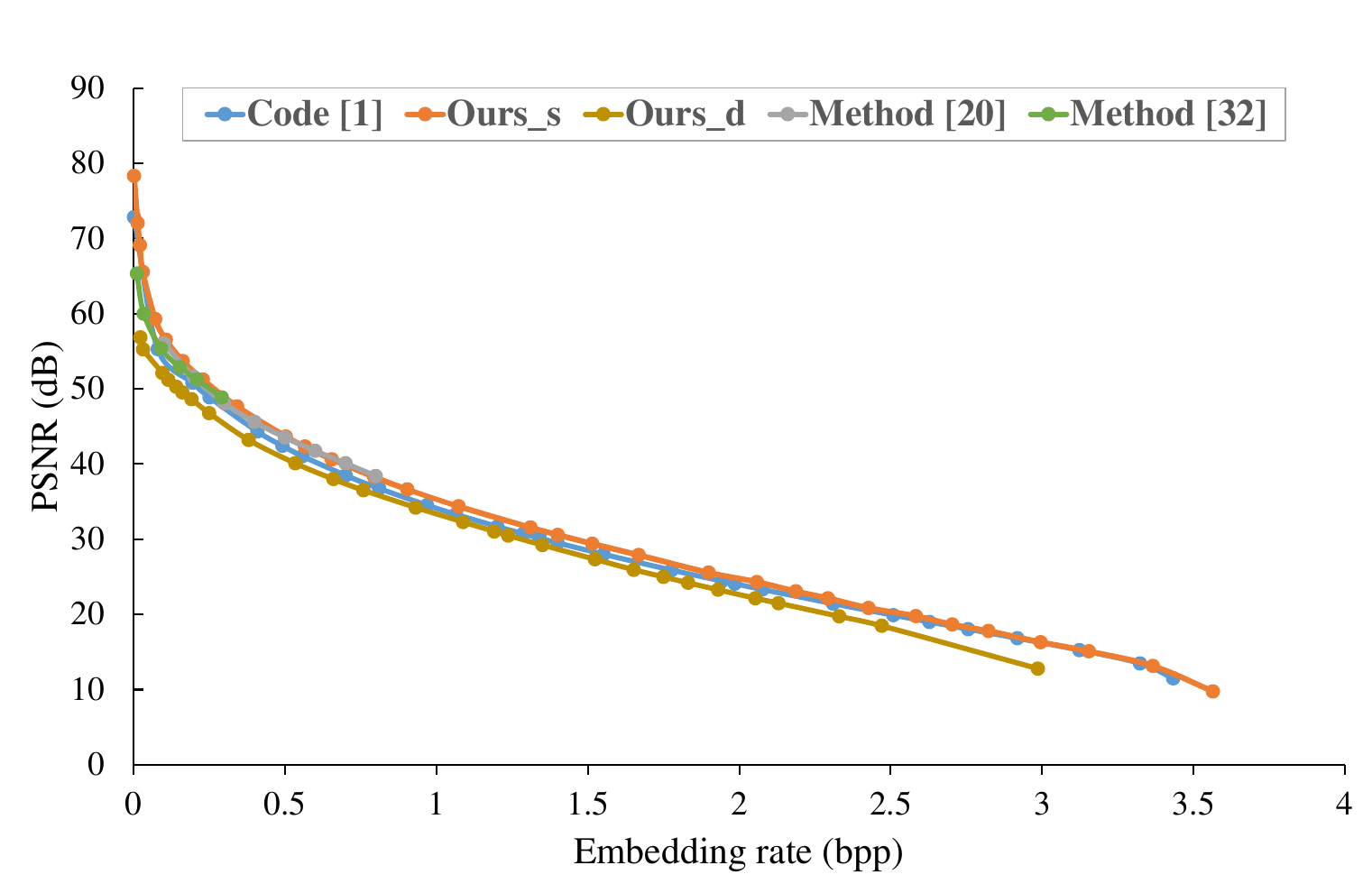}
	\caption{Comparisons in terms of capacity-distortion performance.}
	\label{fig.9}
\end{figure}

Next, we give the experimental results on gray-scale images.
First, Figures \ref{lenatest}--\ref{planetest} shows the embedding rate (bpp) and PSNR (dB) of the proposed static method on two $512\times 512$ sized standard gray-scale images, including Lena and Airplane from the USC\_SIPI image database\footnote{https://sipi.usc.edu/database/}, where the embedding rate $R$ is defined as the number of bits embedded divided by the number of pixels in the host image. The larger $R$, the higher embedding capacity. It can be observed that when the embedded capacity decreases, the PSNR increases steadily.

Second, we show the embedding performance of the proposed static/dynamic method, our previous work \cite{6194314} and two high capacity RDH methods \cite{8733828} and  \cite{JIA2019238} on the test image Lena. We use $32$ bits to represent the state and choose parameters $T=16$, $n=16$ and $v=1$.
The embedding performance with respect to the embedding rate
versus the PSNR curve are plotted in Figure \ref{fig.9}. It shows that for the same embedding capacity, the proposed static method provides slightly higher PSNR values than our previous work~\cite{6194314}. Compared with the method~\cite{8733828}, our static method can achieve the similar performance at the
low embedding rate. However, the method \cite{8733828} states  that it can only produce $0.612$ bpp on average, while the proposed method is capable to achieve larger embedding rates. Further, a location map is required in \cite{8733828} to prevent the underflow/overflow problem, which brings additional storage consumption. Compared with the method~\cite{JIA2019238}, our static proposal can obtain a slight PSNR gain at the
low embedding rate. But, the embedding capacity of the method \cite{JIA2019238} is relatively low, e.g., for the test image Lena, it has an embedding rate of at most $0.29$ bpp. In addition, the method~\cite{JIA2019238} also requires a location map to avoid the underflow/overflow problem. For the proposed dynamic implementation, we initialize the two-dimensional lookup table $\textit{K}$ as follows: $\textit{K}[\hat{h}_i][h_i]=32$ if $\hat{h}_i=h_i$; otherwise, we have $\textit{K}[\hat{h}_i][h_i]=1$. As can be seen in Figure \ref{fig.9}, the proposed dynamic method provides slightly lower PSNR values under the same embedding capacity. This observation is mainly due to the fact that we use the predicted pmf rather than the true pmf. More importantly, unlike our previous work~\cite{6194314}, the proposed dynamic method does not require prior transmission of the host pmf during both embedding and extraction. In addition, compared with the existing methods~\cite{8733828}, \cite{JIA2019238}, our dynamic proposal also allows for larger embedding rates to achieve high capacity requirements.

\section{Conclusion}\label{sec:6}
In this paper, we proposed a novel RDH scheme based on our recent asymmetric numeral systems (ANS) variant~\cite{9810728}. To the best of our knowledge, this paper is the first that employs ANS coding for RDH technique. Unlike our previous work~\cite{6194314}, the proposed method can completely avoid the computing precision problem. In addition, a dynamic implementation is also
adaptively introduced to save the additional overhead of explicitly transmitting the host pmf during embedding and extraction. The
experimental results show that the proposed static method provides slightly higher PSNR values than our previous work~\cite{6194314} and larger embedding rates than some state-of-the-art methods~\cite{8733828}, \cite{JIA2019238} on gray-scale images. Besides, the proposed dynamic method saves additional overhead at the cost of a small image quality loss.

\iffalse
\section{Acknowledgments}\label{sec:7}
This work was supported in part by the National Key Research and Development Program of China under Grant 2022YFA1004902, the National Natural Science Foundation of China under Grant 62071446. The authors thank Ou et al. for sharing the source codes in \cite{8733828}.
\fi

\bibliographystyle{IEEEtran}
\bibliography{ref}

\end{document}